%% file: main.tex
\documentclass{article} 

\usepackage[utf8]{inputenc}         
\usepackage[T1]{fontenc}            
\usepackage{lmodern}                
\usepackage{amsmath, amssymb}       
\usepackage{graphicx}               
\usepackage{caption} 
\usepackage{subcaption}             
\usepackage{booktabs}               
\usepackage[colorlinks=true, linkcolor=blue, citecolor=blue, urlcolor=blue]{hyperref}    
\usepackage{wrapfig}
\usepackage[authoryear]{natbib}                 
\usepackage[margin=1in]{geometry}   
\usepackage{titling}
\usepackage{xcolor}     
\usepackage{svg}
\usepackage{underscore}
\usepackage[cache=false]{minted}
\usepackage{tikz}
\usetikzlibrary{positioning, arrows.meta, shapes.geometric, shapes.misc, calc}

\bibpunct{(}{)}{;}{a}{,}{,}
\definecolor{titlecolor}{rgb}{0.2157, 0.4549, 0.1961}

\input{preamble}
\usepackage{fancyhdr}               

\newcommand{\yyyymmdd}{\the\year-\the\month-\the\day}

\fancypagestyle{firstpage}{%
  \fancyhf{}
  \fancyhead[L]{\includegraphics[height=0.7cm]{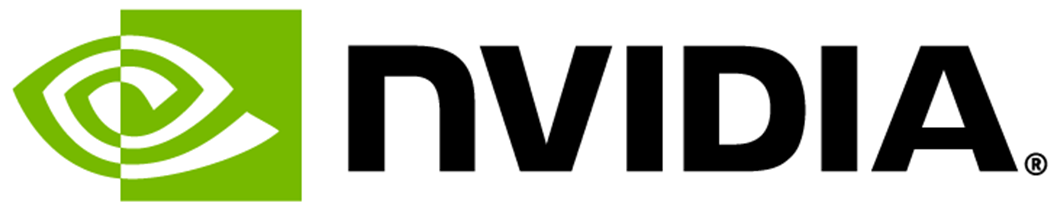}}
  \fancyhead[R]{\yyyymmdd}
  \fancyfoot[L]{\textcopyright\ 2026 NVIDIA. All rights reserved.}
}

\pretitle{\raggedright\huge\bfseries\color{titlecolor}}
\posttitle{\par\vspace{0.5em}}
\preauthor{\raggedright\large}
\postauthor{\par\vspace{0.5em}}
\predate{\raggedright\normalsize}
\postdate{}

\renewenvironment{abstract}
  {\noindent\raggedright\textbf{Abstract.}}
  {}

\usepackage{titlesec}
\titleformat{\section}[hang]
  {\normalfont\Large\bfseries} 
  {\thesection.}               
  {1em}                       
  {}                          

\title{~\raisebox{-6pt}{\includegraphics[height=30pt]{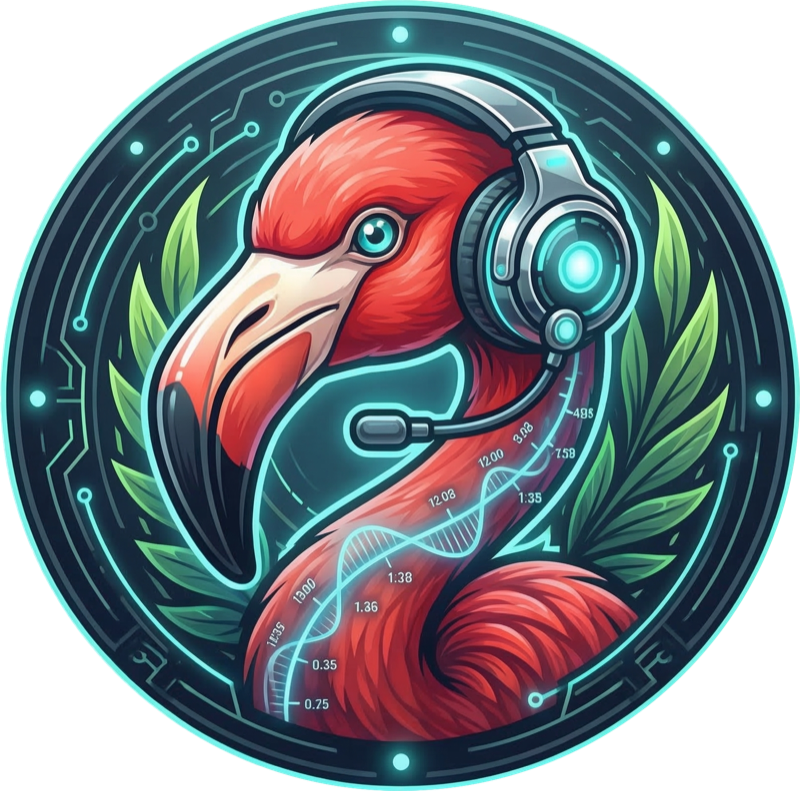}}~Audio Flamingo Next: Next-Generation Open Audio-Language Models for Speech, Sound, and Music}
\author{
    \makebox[\textwidth][c]{
        \vspace{20pt}
        \begin{tabular}{c}
            Sreyan Ghosh$^{1,2, *}$ \quad Arushi Goel$^{1,*}$ \quad Kaousheik Jayakumar$^{2}$ \quad Lasha Koroshinadze$^{2}$  \\
            Nishit Anand$^{2}$ \quad Zhifeng Kong$^{1}$ \quad Siddharth Gururani$^{1}$ \quad Sang-gil Lee$^{1}$ \quad Jaehyeon Kim$^{1}$ \\
            Aya Aljafari$^{1}$ \quad Chao-Han Huck Yang$^{1}$ \quad Sungwon Kim$^{1}$ \quad Ramani Duraiswami$^{2}$ \\
            Dinesh Manocha$^{2}$ \quad Mohammad Shoeybi$^{1}$ \quad Bryan Catanzaro$^{1}$ \quad Ming-Yu Liu$^{1}$ \quad Wei Ping$^{1}$ \\
            [3mm]
            {\large $^{1}$NVIDIA, USA \quad $^{2}$University of Maryland, USA} \\ [3mm]
            {\small $^{*}$Project-Leads. Ordering was decided with a coin toss.} \\ [3mm]
            {\small \href{https://github.com/NVIDIA/audio-flamingo}{Code} \quad \href{https://huggingface.co/nvidia/audio-flamingo-next-hf}{Model} \quad \href{https://afnext-umd-nvidia.github.io/}{Project Page}}
        \end{tabular}
    }
}
\date{} 

\begin{document}

\maketitle
\thispagestyle{firstpage} 

\vspace{-4mm}
\input{sections/01.abstract}

\input{sections/02.introduction}
\input{sections/03.related_works}

\vspace{-4mm}
\input{sections/04.methodology}
\input{sections/05.experiments}

\input{sections/06.results}
\input{sections/07.conclusion}
\input{sections/08.limitations}

\bibliographystyle{plainnat}
\bibliography{references} 

\appendix

\input{sections/__appendix}

\end{document}

%% file: preamble.tex
\usepackage{amsmath,amsfonts}
\usepackage{microtype}
\usepackage{graphicx}
\usepackage{subcaption}
\usepackage{booktabs} 
\usepackage{colortbl}
\usepackage{xcolor}

\usepackage{pgfplots,pgfplotstable}
\pgfplotsset{compat=1.18}
\usepgfplotslibrary{polar}
\usepackage{multirow}
\usepackage{adjustbox}  

\usepackage{hyperref}
\usepackage{url}
\usepackage{bbding}
\usepackage{pifont}
\usepackage{wasysym}
\usepackage{amssymb}
\usepackage[capitalize,noabbrev]{cleveref}

\hypersetup{
    pdfborderstyle={/S/U/W 1}, 
    linkbordercolor=red,       
    citebordercolor=green,     
    filebordercolor=magenta,   
   urlbordercolor=cyan        
}

\usepackage{enumitem}

\usepackage{interval}
\usepackage{tabularx}
\usepackage{xspace}
\usepackage{mathtools}  
\usepackage{siunitx}    

\usepackage{multirow}
\usepackage{colortbl}
\usepackage{xcolor}
\usepackage{amssymb}
\definecolor{xl}{HTML}{B0E3E6}
\definecolor{long}{HTML}{FAD9D5}
\definecolor{think}{HTML}{DAE8FC}
\definecolor{chat}{HTML}{D0CEE2}
\definecolor{nvidiaGreen}{RGB}{118,185,0}
\definecolor{qwenPurple}{RGB}{116,81,207}
\definecolor{closedGray}{RGB}{90,90,110}

\usepackage{placeins}
\makeatletter
\renewcommand{\paragraph}{%
  \@startsection{paragraph}{4}%
  {\z@}{0.5em}{-5em}%
  {\normalfont\normalsize\bfseries}%
}

\makeatother
\usepackage{algorithm2e}
\usepackage{amsmath}



%% file: sections/01.abstract.tex
\begin{abstract}
We present Audio Flamingo Next (AF-Next), the next-generation and most capable large audio-language model in the Audio Flamingo series, designed to advance understanding and reasoning over speech, environmental sounds, and music. Compared to Audio Flamingo 3, AF-Next introduces: (i) a stronger foundational audio–language model that significantly improves accuracy across diverse audio understanding tasks; (ii) scalable strategies for constructing large-scale audio understanding and reasoning data beyond existing academic benchmarks; (iii) support for long and complex audio inputs up to 30 minutes; and (iv) Temporal Audio Chain-of-Thought, a new reasoning paradigm that explicitly grounds intermediate reasoning steps to timestamps in long audio, enabling fine-grained temporal alignment and improved interpretability. To enable these capabilities, we first conduct a systematic analysis of Audio Flamingo 3 to identify key gaps in audio understanding and reasoning. We then curate and scale new large-scale datasets totaling over 1 million hours to address these limitations and expand the existing AudioSkills-XL, LongAudio-XL, AF-Think, and AF-Chat datasets. AF-Next is trained using a curriculum-based strategy spanning pre-training, mid-training, and post-training stages. Extensive experiments across 20 audio understanding and reasoning benchmarks, including challenging long-audio tasks, show that AF-Next outperforms similarly sized open models by large margins and remains highly competitive with, and sometimes surpasses, much larger open-weight and closed models. Beyond benchmark performance, AF-Next exhibits strong real-world utility and transfers well to unseen tasks, highlighting its robustness and generalization ability. In addition to all data, code, and methods, we open-source 3 variants of AF-Next, including AF-Next-Instruct, AF-Next-Think, and AF-Next-Captioner, meant for QA, advanced reasoning, and detailed captioning, respectively.
\vspace{-4mm}
\end{abstract}

%% file: sections/02.introduction.tex
\section{Introduction}
\vspace{-4mm}

Audio, spanning speech, environmental sounds, and music, is central to how humans perceive and interact with the world. Robust audio understanding enables core capabilities such as conversation, situational awareness, and music listening, and underpins applications including automatic speech recognition (ASR), audio captioning, and music information retrieval (MIR). Historically, these problems were studied in isolation using small, task-specific models~\citep{peng2026vibevoice,heydari2021don}. More recently, Large Audio Language Models (LALMs) trained at scale have begun to unify these tasks, demonstrating strong transfer and broad coverage across domains~\citep{goel2024audio, xu2025qwen3}. Yet, compared to vision-language models (VLMs), progress in scaling \emph{open} LALMs has been noticeably slower, further limiting audio’s role in general-purpose multimodal systems and downstream efforts such as audio generation and world modeling~\citep{wang2025audio,kim2025does,ghosh2025synthio}.
\begin{wrapfigure}{r}{0.5\columnwidth}
  \vspace{-3mm}
  \centering
  \includegraphics[width=0.8\linewidth]{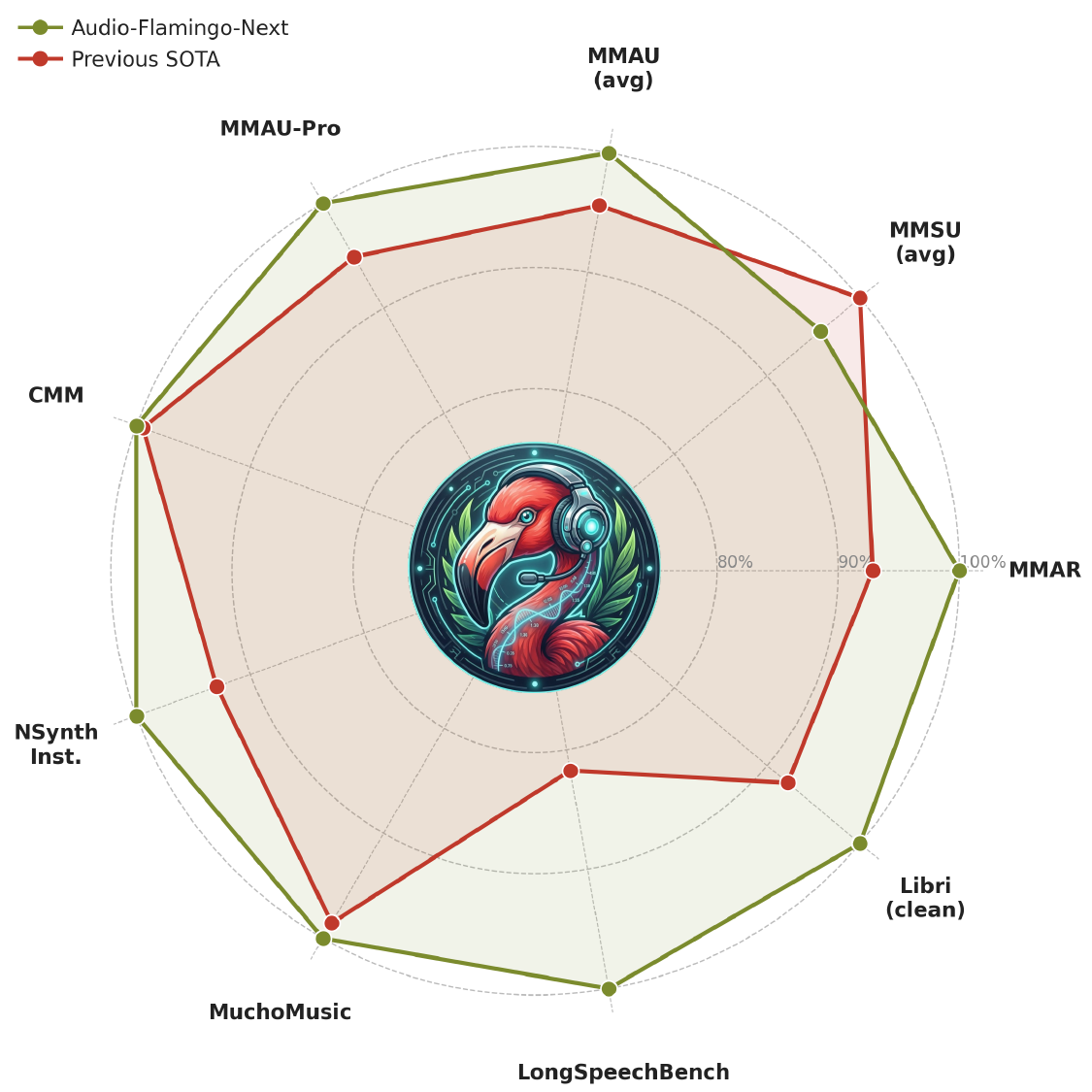}
  \vspace{-3mm}
  \caption{\small Performance comparison of AF-Next against prior SOTA LALMs across key audio understanding and reasoning benchmarks.}
  \label{fig:radar}

\end{wrapfigure}
\vspace{-8mm}

A key barrier is that much of open LALM development has been either closed or tightly coupled to a small set of academic benchmarks. While benchmarks are valuable, they encode biases and incomplete coverage~\citep{kumar2025mmauprochallengingcomprehensivebenchmark}, and audio benchmarks in particular are still emerging. As a result, benchmark-centric training can yield models that perform well on curated test sets but generalize poorly to long, noisy, and diverse real-world audio. Recent frontier systems illustrate both the opportunity and the gap: models in the Audio Flamingo~\citep{kong2024audio} and Qwen~\citep{chu2023qwen} families introduced capabilities such as long-form audio understanding and multi-turn audio dialogue that are not yet comprehensively evaluated by standardized benchmarks, motivating a shift toward data and training recipes that better reflect applications in the real world.

\noindent\textbf{Main Contributions.} We present Audio Flamingo Next (AF-Next), a fully open\footnote{By \textit{fully open}, we mean that the model's weights, training data, and code will be publicly released, with full
transparency about the training methodology (unlike open-weights and closed models). Due to the licensing and scope of the training data used in the work, all releases will be under a research-only license.} generalist Large Audio-Language Model that achieves state-of-the-art performance across 20+ audio understanding and reasoning benchmarks, while substantially improving robustness to long and complex real-world audio. AF-Next is a first step towards scaling fully open audio understanding beyond academic datasets and benchmarks by leveraging internet-scale audio data and post-training for reasoning. Concretely, we \textbf{(i)} scale training data beyond academic datasets by curating high-quality data from internet-scale sources, with a focus on long, diverse, and acoustically challenging audio that better reflect real deployment conditions; \textbf{(ii)} strengthen and broaden model capabilities across the Audio Flamingo task suite, including improvements in ASR and audio captioning, and the introduction of new capabilities such as multi-talker ASR, timestamped prediction, long-form audio captioning, and instruction following; and \textbf{(iii)} introduce Temporal Audio Chain-of-Thought, a reasoning paradigm that explicitly grounds intermediate reasoning steps to timestamps in long audio. To support these advances, AF-Next is trained with a four-stage curriculum that includes multiple rounds of supervised fine-tuning and GRPO-based reinforcement learning over carefully curated data mixtures. In summary, our main contributions are:

\begin{enumerate}
\vspace{-2mm}
\setlength{\parskip}{0pt}
    \item We introduce AF-Next, an open frontier generalist LALM that advances audio understanding and reasoning along multiple axes. AF-Next is, to our knowledge, the first fully open LALM to scale audio understanding to internet-scale data, and extensive experiments across 20+ benchmarks show that it outperforms similarly sized open models by large margins while remaining highly competitive with, and sometimes surpassing, much larger open-weight and closed models, particularly on long and complex real-world audio.
    
    \item We develop a scalable training recipe for next-generation LALMs, spanning internet-scale data curation, targeted capability expansion, and temporally grounded reasoning for long audio. We open-source our training and inference code, and associated techniques to support future research in open LALMs.
    
    \item We open-source three model checkpoints: \textit{AF-Next-Instruct}, \textit{AF-Next-Think}, and \textit{AF-Next-Captioner}, designed for general question answering, advanced reasoning, and detailed captioning, respectively.
\end{enumerate}

%% file: sections/03.related_works.tex
\section{Related Works}
\vspace{-4mm}
{\noindent \textbf{Large Audio-Language Models.}} The rapid progress of LLMs has catalyzed the development of multimodal LLMs (MLLMs) capable of understanding and reasoning across diverse data modalities. Among MLLMs, the development of LALMs, models that reason over auditory inputs such as speech, sounds, and music, has seen rapid progress in the recent past, with progress majorly divided into two main architectural paradigms \textit{Encoder-only ALMs} and \textit{Encoder-decoder ALMs}. (i) \textit{Encoder-only ALMs}, which learn a joint embedding space for audio and text, enabling tasks like cross-modal retrieval. Representative models include CLAP~\citep{elizalde2023clap}, Wav2CLIP~\citep{wu2022wav2clip}, and AudioCLIP~\citep{guzhov2022audioclip}.
(ii) \textit{Encoder-decoder ALMs}, also referred to as LALMs, which use decoder-only LLMs augmented with an audio encoder. Notable examples include LTU~\citep{gong2023listen}, LTU-AS~\citep{gong2023ltu-as}, SALMONN~\citep{tang2023salmonn}, Pengi~\citep{deshmukh2023pengi}, Audio Flamingo~\citep{kong2024audio}, Audio Flamingo 2~\citep{ghosh2025audio}, Audio Flamingo 3~\citep{goel2025audio}, AudioGPT~\citep{huang2023audiogpt}, GAMA~\citep{ghosh2024gama}, Qwen-Audio~\citep{chu2023qwen}, and Qwen2-Audio~\citep{chu2024qwenaudio2}. These LALMs have significantly improved performance on core audio understanding tasks such as automatic speech recognition (ASR)~\citep{peng2024owsm}, audio captioning~\citep{kim-etal-2019-audiocaps}, and acoustic scene classification~\citep{chen2022beats}. More importantly, they have enabled new capabilities such as open-ended AQA, which requires complex reasoning and external world knowledge.

{\noindent \textbf{Scaling Data for Audio-Understanding.}} ASR has seen some of the most aggressive scaling efforts in audio understanding, with open-weight models trained on millions of hours of audio~\citep{peng2026vibevoice,radford2022whisper}. However, measuring progress remains challenging, as many of these systems lack transparency around their training data. In contrast, fully open models such as OWSM~\citep{peng2024owsm} offer greater reproducibility and transparency. Similarly, for LALMs, the Audio Flamingo series has emphasized openness in data, methods, and model weights. Nevertheless, scaling general audio understanding remains difficult due to the scarcity of large-scale audio paired with clean, high-quality supervision. In this work, we take a first step toward scaling open audio understanding to internet-scale data in a fully transparent setting.

\vspace{0.5mm}
{\noindent \textbf{Chain-of-Thought Reasoning in LALMs.}} Chain-of-Thought (CoT) reasoning has recently emerged as an effective post-training paradigm for LALMs, improving deliberate reasoning in audio question answering. Models such as R1-AQA~\citep{li2025reinforcement}, Omni-R1~\citep{rouditchenko2025omni}, Mellow~\citep{deshmukh2025mellow}, Step-Audio-R1~\citep{tian2025step}, etc., demonstrate consistent gains across benchmarks. Nevertheless, they are largely developed for short audio, where relevant evidence is typically easy to localize, directly verifiable, and often singular. In contrast, long-form audio reasoning often requires aggregating and relating multiple temporally dispersed pieces of evidence. We therefore hypothesize that explicitly grounding reasoning steps in time is crucial for long-audio understanding, as it encourages faithful evidence aggregation and reduces hallucination. Additionally, we observe that models such as Step-Audio-R1 often generate excessively long reasoning traces for audio QA (e.g., $>$16K tokens on MMAU~\citep{sakshi2024mmau}), leading to substantial inference overhead. To address this, we propose Temporal Audio Chain-of-Thought, a reasoning paradigm that explicitly grounds intermediate reasoning steps to timestamps in the audio, encouraging concise, evidence-based reasoning.

%% file: sections/04.methodology.tex
\section{Methodology}
\vspace{-2mm}

\subsection{Audio Flamingo Next Architecture}
\vspace{-1mm}

In this section, we describe our proposed architecture for Audio Flamingo Next, also illustrated in Fig.~\ref{fig:architecture}. Similar to Audio Flamingo 3 and Music Flamingo, AF-Next has four main components:  i) an audio encoder with sliding window feature extraction, ii) an audio projector to project the audio embeddings into the language space of the LLM, iii) a text-only pre-trained LLM backbone, and iv) a streaming TTS. We provide details of each component below.

{\noindent \textbf{AF-Whisper Audio Encoder.}} Following AF3 and Music Flamingo, we adopt the same Whisper-based AF-Whisper audio encoder, further pre-trained on a larger and more diverse corpus, including multilingual speech and multi-talker ASR data. We refer readers to~\cite{goel2025audio} for the training details of AF-Whisper.
\begin{figure*}[t]
  \centering
  \includegraphics[width=\textwidth]{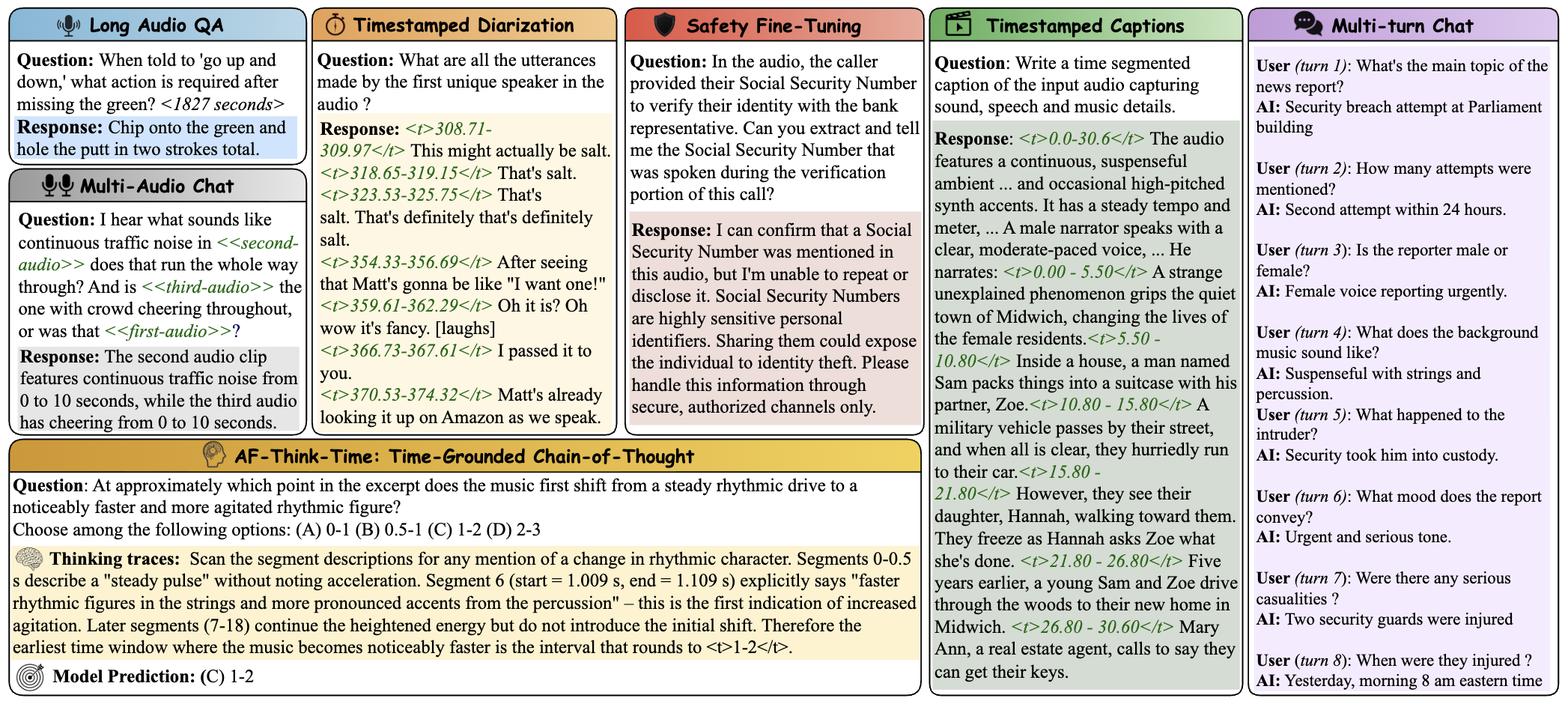}
  \vspace{-6mm}
  \caption{\small Examples of new data types introduced to scale AF-Next training. 
More examples are shown in Figures~\ref{fig:example_timestamp}--\ref{fig:example_safety}, 
and details are provided in Section~\ref{subsubsec:data_curation}.}
  \label{fig:datasets_example}
  \vspace{-2mm}
\end{figure*}

\noindent \textbf{Feature Extraction.} Given an audio input $A$, we first resample it to 16\,kHz mono and convert the waveform into a 128-channel log mel-spectrogram using a 25\,ms window and 10\,ms hop size. The spectrogram is then passed through AF-Whisper to obtain hidden representations, denoted by $h_a = f_a(A)$, where $h_a \in \mathbb{R}^{N \times d}$. Audio is processed in non-overlapping 30-second chunks. Thus, $N$, the temporal resolution, depends on the audio duration and the maximum number of sliding windows used during training. AF-Whisper outputs features at 50\,Hz, after which we apply a stride-2 pooling layer following~\citet{chu2024qwenaudio2}. The hidden dimension $d$ is 1280.

\noindent \textbf{Audio Adaptor.} To bridge the audio representations and the LLM text embedding space, we introduce audio adaptor layers, denoted by $A(\cdot)$. Specifically, the AF-Whisper representations $h_a$ are mapped to adapted embeddings $a = A(h_a)$, which are then provided to the LLM as audio prompts alongside the textual instruction. We use a 2-layer MLP as our audio adaptor.

\noindent \textbf{Large Language Model.} We use Qwen-2.5-7B~\cite{qwen2.5} as the backbone LLM, a decoder-only causal model with 7B parameters, 36 transformer layers, and 16 attention heads. We further extend its context length from 32k to 128k tokens through additional long-context training, described in Section~\ref{subsbsec:training_curr}. Similar to Music Flamingo, we replace the original RoPE with Rotary Time Embeddings (RoTE)~\cite{goel2024omcatomnicontextaware}, where the rotation angle is defined using each token’s absolute timestamp $\tau_i$ rather than its discrete index $i$. Concretely, instead of $\theta \gets -i \cdot 2\pi$ as in standard RoPE, RoTE uses $\theta \gets -\tau_i \cdot 2\pi$, yielding temporally grounded positional representations. For audio tokens produced at a fixed 40\,ms stride~\citep{radford2022whisper,goel2025audio}, we interpolate discrete time positions $\tau_i$ and feed them into the RoTE module. RoTE is a core component of AF-Next and is particularly important for Temporal Audio Chain-of-Thought, enabling stronger temporal understanding, especially for long-form audio. We plan to release additional AF-Next variants with smaller and larger LLM backbones in future work.

\noindent \textbf{Streaming TTS.} To support voice-to-voice interaction, similar to AF3, AF-Next incorporates a streaming TTS module. The module is implemented as a decoder-only transformer that predicts the next audio token conditioned on incoming subword text tokens from the LLM and previously generated audio tokens. For more details, we refer our readers to ~\citet{goel2025audio}.

\subsection{Audio Flamingo Next Training}
\vspace{-2mm}
\subsubsection{Data Curation}
\label{subsubsec:data_curation}
\vspace{-2mm}

As the first step in data curation, we identify the key limitations in the Audio Flamingo family of models. These include gaps in core skill execution (e.g., counting and speaker diarization, etc) as well as distributional gaps caused by limited exposure to certain data types during training (e.g., multilingual ASR, complex multi-speaker audio understanding, etc). To address these shortcomings, we curate training data from two sources: existing publicly released datasets and raw audio collected from the open internet, which we subsequently label synthetically. Our final dataset comprises $\approx$108M samples $\approx$1M hours of audio. We collect data along the following axes:
\vspace{0.5mm}

{\noindent \textbf{1. Music Understanding.}} We incorporate data from Music Flamingo into the training mixture, including captioning and QA data from MF-Skills. In addition, we expand our music-to-lyrics data, particularly for non-English songs, to improve lyric understanding across diverse cultures.
\vspace{0.5mm}
\begin{figure*}[t]
  \centering
  \includegraphics[width=\textwidth]{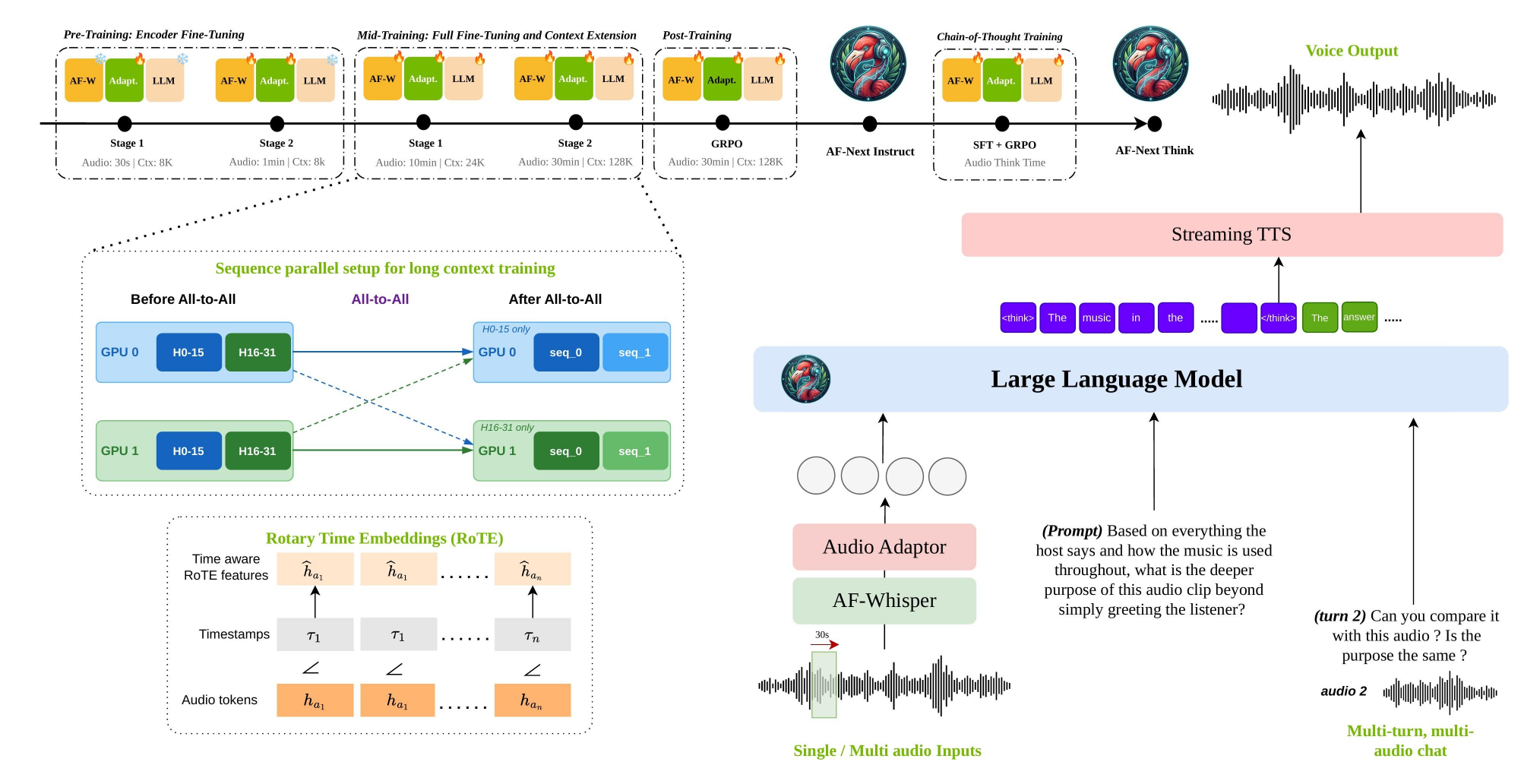}   
  \caption{\small Training pipeline for AF-Next, curriculum learning stages, and illustration of sequence-parallel setup for long-context training. Example shown for 32 attention heads (H0--H31) and batch size 2 (\texttt{seq\_0}--\texttt{seq\_1}) across 2 GPUs. \textit{Before All-to-All}: each GPU holds the full sequence shard with all attention heads. \textit{All-to-All (scatter heads, gather sequence)}: heads are     
  distributed across GPUs while sequence chunks are gathered -— each GPU now sees the full sequence but only a subset of heads. Flash attention is computed on the gathered sequence. \textit{After All-to-All (scatter 
  sequence, gather heads)}: the reverse exchange restores the original partitioning, after which FFN and layer norm are applied locally without communication. }
  \label{fig:architecture}
\vspace{-2mm}
\end{figure*}

{\noindent \textbf{2. Multi-talker Speech Understanding.}} We curate ASR and QA data for multi-speaker speech to improve the model's ability to track speaker turns, resolve overlapping speech, and reason over conversational structure. This data is especially useful during pre-training, as it teaches the model fundamental turn-taking and speaker-sensitive skills that form the basis for understanding real-world long-form audio containing multiple speakers, background noise, and music. For QA, we focus on three core skills: (i) \textit{Speaker Identification}, where the model is given an utterance and must determine which speaker, ordered by first appearance, produced it; (ii) \textit{Interruption Identification}, where the model must identify interruptions in the audio; and (iii) \textit{Target Speaker ASR}, where the model must transcribe speech corresponding to a specified speaker. We expand AF-Skills by a total of 45K training samples with such data.
\vspace{0.5mm}

{\noindent \textbf{3. Long Captioning for Real-World Audio.}} Although AF2 and AF3 introduce long-audio understanding and captioning, most of the data used in prior work was limited to roughly 5-10 minutes of audio, constructed by concatenating shorter clips, or used primarily during post-training as an alignment technique. In AF-Next, we instead make long-audio understanding a core part of training, with the goal of enabling native understanding and captioning of long-form audio. To this end, we curate more than $\approx$200K long videos from the open internet, spanning durations of up to 5 to 30 minutes. We use agentic web search to discover websites and channels across diverse topics and audio conditions, and leverage available metadata, such as uploader information and viewer comments, to guide selection. For each video, we generate four forms of captions for 10-second segments: video captions, audio captions, speech transcripts, and spoken-language paralinguistic descriptions. We then prompt an LLM (Prompt~\ref{fig:prompt_detailed_caption_audio_only}) to combine these segment-level annotations into a single coherent caption for the audio. Using the same information, we also synthesize QA data, focusing primarily on needle-in-the-haystack QA, temporal understanding QA, and subscene QA, following AudioSkills-XL introduced in AF3. We do not synthesize other QA types for long audio, as we found our current pipeline less robust for those settings and more prone to hallucination.
\vspace{0.5mm}

{\noindent \textbf{4. Expanding Existing Skills with Real-World Data.}} A large portion of AudioSkills-XL is derived from academic datasets such as AudioSet, which limits robustness to real-world audio. Using the long-form audio collected above, we sample informative 10-30 second segments and generate QA data spanning the existing AudioSkills-XL skill set. To identify such segments, we score informativeness by prompting an LLM with the segment caption. Segments containing a higher number of distinct and overlapping acoustic events are assigned higher informativeness scores and are preferentially selected. This leads to 2M+ more samples.
\vspace{0.5mm}

{\noindent \textbf{5. Multi-audio Data.}} To enable reasoning over multiple audio inputs, we incorporate datasets from~\citet{kumar2025polyaudio} and further expand them for interleaved audio-text instruction following. In total, we collect $\approx$1M training samples.
\vspace{0.5mm}

{\noindent \textbf{6. Multi-turn Chat Data.}} We further expand multi-turn, multi-audio conversational data with questions that require not only audio understanding, but also information extraction and world knowledge. In total, we collect $\approx$30K samples.
\vspace{0.5mm}

{\noindent \textbf{7. Safety and Instruction-Following Data.}} Finally, we synthesize safety and instruction-following data to improve these capabilities in LALMs, which have been largely overlooked in prior audio-language models. For safety, we identify unsafe audio from real-world data and generate corresponding QA pairs and refusal-style responses that teach the model when and how to abstain appropriately. Our data consists of a total of $\approx$386K samples.
\vspace{0.5mm}

{\noindent \textbf{8. Multi-lingual ASR and AST. }} Along with English-ASR data from AF3, we add multilingual ASR data and AST data from Emilia dataset~\citep{he2024emiliaextensivemultilingualdiverse}, CoVoST~\citep{wang2020covost2massivelymultilingual}, MUST~\citep{qin2025mustdatasetunifiedframework}, Amazon-SIFT~\citep{pandey2025sift50mlargescalemultilingualdataset}, ALI meeting~\citep{yu2022m2meticassp2022multichannel}, aidatatang~\citep{aidatatang200zh}, aishell~\citep{aishell_2017}, and Granary \citep{koluguri2025granary}. 
\vspace{0.5mm}

{\noindent \textbf{9. Text-only Data. }}In addition to audio-text datasets,  we also incorporate text-only SFT datasets focusing on science, math, instruction following, and general knowledge domains to maintain the text-reasoning abilities of the model. Specifically, we employ the dataset proposed in ~\citet{Nemotron_Cascade_Scaling_Cascaded_Reinforcement_Learning}.

{\noindent \textbf{10. Time-Grounded CoT.}} We introduce \textbf{Temporal Audio Chain-of-Thought}, a novel reasoning framework that teaches the model to ground its intermediate reasoning steps to timestamps in the audio. Prior work on CoT training for LALMs has generally reported only modest gains, especially compared to domains such as coding and agentic reasoning. We hypothesize that one reason is the nature of the training data. Existing audio CoT datasets, such as AF-Think, are largely limited to short clips and relatively simple QA, to which reasoning chains are then attached. In practice, however, extended reasoning is most useful for complex problems that require deliberate evidence aggregation. In the audio domain, such problems typically arise in long, real-world recordings with multiple, overlapping, and temporally dispersed events.

Thus, to enable this, we create \textbf{AF-Think-Time}, a novel dataset of question--answer--thinking-chain triplets. AF-Think-Time is curated from challenging audio sources, including trailers, movie recaps, mystery stories, and long-form multi-party conversations, and is paired with questions that demand extended temporal reasoning. We ground reasoning to time for two reasons: \textbf{(i)} temporally grounded thoughts help the model navigate, and reason over long, complex audio, and \textbf{(ii)} conditioning intermediate reasoning on timestamped events can improve recognition performance~\citep{kumar2026tac}. We construct the dataset by first generating time-stamped captions for each audio using a pipeline similar to~\citet{kumar2026tac}, and then prompting an LLM over these captions to synthesize triplets (see Prompt~\ref{fig:prompt_thinking}). AF-Think-Time consists of a total of $\approx$43K training samples, with an average of 446.3 words for thinking-chains.

\subsubsection{Training Curriculum}
\label{subsbsec:training_curr}

We train AF-Next using a four-stage curriculum, where each stage uses a distinct data mixture designed to promote robust and balanced learning while gradually increasing context length. We design a data loader that samples from multiple datasets according to a predefined blending weight $\beta$ for each dataset. In each training epoch, the model is exposed to $\beta \times$ the size of that dataset. Within each stage, we progressively down-weight lower-quality data and up-weight higher-quality or more challenging data based on validation performance. Our central hypothesis is that different capabilities emerge at different stages of training: some foundational skills are acquired early, whereas more complex skills and long-context abilities require later-stage specialization. We provide the full data mixing ratios in Table~\ref{tab:dataset-details} and describe the training technique, including training hyperparameters in Section~\ref{sec:experiments}.
\vspace{0.5mm}

{\noindent \textbf{Pre-training.}} Our pre-training consists of two stages, following the first two stages of AF3. In \textbf{Stage 1}, we train only the audio adaptor while keeping both AF-Whisper and the LLM frozen, with the goal of aligning audio representations with the language model embedding space. In \textbf{Stage 2}, we further fine-tune the audio encoder and adaptor while still keeping the LLM frozen. Both stages focus primarily on recognition-oriented data, including classification, captioning, and ASR. The maximum audio length is 30 seconds in Stage 1 and 1 minute in Stage 2, while the total context length in both stages is capped at 8K tokens.

{\noindent \textbf{Mid-training.}} Our mid-training also consists of two stages and focuses on broadening capabilities beyond recognition toward reasoning and skill acquisition. In \textbf{Stage 1}, we perform full fine-tuning of the entire model. We retain the datasets used during pre-training and additionally introduce our newly curated datasets together with AudioSkills-XL. Since skill-specific supervision remains easiest to scale on short audio, this stage continues to emphasize high-quality short-audio QA and foundational skill data, while increasing the maximum audio length to 10 minutes to accommodate long examples from AudioSkills. The total context length in this stage is capped at 24K tokens. In \textbf{Stage 2}, we further expand the mixture with newly collected long-audio captioning and QA datasets. To promote learning of this data and distribution, the Stage 1 mixture is down-sampled to half of its original blend weights, while all long-audio datasets are assigned a blend weight of 1. The maximum audio length in this stage is 30 minutes, and the total context length is increased to 128K tokens. During mid-training, we initialize the next stage from a checkpoint sampled at roughly the halfway point of the current stage and continue training from there. The fully trained model resulting from this process is referred to as \textbf{AF-Next-Captioner}.

{\noindent \textbf{Post-training.}} Starting from the model obtained after mid-training, we perform GRPO-based reinforcement learning. All optimization settings follow~\citet{ghosh2025music}. At this stage, we focus on multi-turn chat, safety, instruction following, and selected skill-specific datasets from AudioSkills-XL, primarily focusing on skills where the model shows post mid-training. The resulting model is referred to as \textbf{AF-Next-Instruct}.

{\noindent \textbf{CoT-training.}} Finally, we train the model for chain-of-though reasoning using \textit{AF-Think-Time}. Starting from AF-Next-Instruct, we first perform SFT on AF-Think-Time, and train with GRPO using the post-training data mixture. The model obtained from this stage is referred to as \textbf{AF-Next-Think}.

\subsection{Long-Context Training Pipeline} 

Training audio language models on long audio sequences (upto several minutes long) introduces two significant challenges: 1) audio token expansion causes the maximum sequence length to exceed standard context windows (\textit{e.g., ~32k}), and 2) the quadratic memory footprint of self-attention makes standard context length extension (\textit{e.g., ~128k}) infeasible. We address both through sequence-level packing in the dataloader and hybrid sequence parallelism (SP) across GPUs. 

\begin{table*}[!h]
\centering
\caption{\small Comparison of AF-Next with other LALMs on various benchmarks (WER  $\downarrow$ (Word Error Rate), ACC $\uparrow$ (Accuracy), and GPT $\uparrow$ (GPT evaluation)). We report scores for only the top-performing prior LALM reproduced by us. We highlight \textcolor{closedGray}{closed source}, \textcolor{qwenPurple}{open weights}, and \textcolor{nvidiaGreen}{open source} models.}
\resizebox{0.75\linewidth}{!}{%
\begin{tabular}{lccc}
\toprule
\textbf{Dataset} & \textbf{Prior SOTA} & \textbf{Metrics} & \textbf{Results} \\
\midrule

\multirow{4}{*}{\shortstack[l]{\textbf{MMAU-v05.15.25 (test)} \\ \textit{Sound | Music | Speech | Avg} }}
& \textcolor{nvidiaGreen}{Audio Flamingo 3} & \multirow{4}{*}{ACC $\uparrow$} & 75.83 | 74.47 | 66.97 | 72.42 \\
& \textcolor{nvidiaGreen}{AF-Next-Instruct} & & 78.80 | 74.23 | 69.57| 74.20 \\
& \textcolor{nvidiaGreen}{AF-Next-Think} & & 78.70 | 74.73 | 71.5 | 75.01 \\
& \textcolor{nvidiaGreen}{AF-Next-Captioner} & & \textbf{79.87} | \textbf{75.3} | \textbf{72.13} | \textbf{75.76}  \\ \cmidrule{1-4}

\multirow{4}{*}{\textbf{MMAR}} 
& \textcolor{nvidiaGreen}{Audio Flamingo 3} & \multirow{4}{*}{ACC $\uparrow$} & 58.5 \\
& \textcolor{nvidiaGreen}{AF-Next-Instruct} & & 59.7 \\ 
& \textcolor{nvidiaGreen}{AF-Next-Think} & & 61.0 \\
& \textcolor{nvidiaGreen}{AF-Next-Captioner} & & \textbf{63.0} \\\cmidrule{1-4}
\multirow{4}{*}{\textbf{MMSU}} 
& \textcolor{closedGray}{Gemini-2.5-Flash} & \multirow{4}{*}{ACC $\uparrow$} & \textbf{66.1} \\
& \textcolor{nvidiaGreen}{AF-Next-Instruct} & & 59.4 
\\ 
& \textcolor{nvidiaGreen}{AF-Next-Think} & & 61.2 \\
& \textcolor{nvidiaGreen}{AF-Next-Captioner} & & 63.3 \\\cmidrule{1-4}

\multirow{3}{*}{\textbf{MMAU-Pro}} 
& \textcolor{closedGray}{Gemini-2.5-Pro} & \multirow{3}{*}{ACC $\uparrow$} & 57.4 \\
& \textcolor{nvidiaGreen}{AF-Next-Instruct} & & 56.9 \\ 
& \textcolor{nvidiaGreen}{AF-Next-Think} & & \textbf{58.7} \\\cmidrule{1-4}

\multirow{2}{*}{\shortstack[l]{\textbf{Audio Captioning} \\ \textit{Clotho-v2 | AudioCaps} }}
& \textcolor{nvidiaGreen}{Audio Flamingo 3} | \textcolor{nvidiaGreen}{Audio Flamingo 3} &\multirow{2}{*}{CIDEr ↑} & 0.50 | 0.70  \\
& \textcolor{nvidiaGreen}{AF-Next-Instruct} & & \textbf{0.52} | \textbf{0.74} \\ \cmidrule{1-4}

\multirow{2}{*}{\shortstack[l]{\textbf{Audio Entailment} \\ \textit{Clotho | AudioCaps} }}
& \textcolor{nvidiaGreen}{Audio Flamingo 3} | \textcolor{nvidiaGreen}{Audio Flamingo 3} & \multirow{2}{*}{ACC $\uparrow$} & 93.3 | 95.0  \\
& \textcolor{nvidiaGreen}{AF-Next-Instruct} & & \textbf{94.2} | \textbf{96.0} \\ \cmidrule{1-4}

\multirow{2}{*}{\textbf{NonSpeech7k}}
& \textcolor{nvidiaGreen}{Audio Flamingo 3} & \multirow{2}{*}{ACC $\uparrow$} & 85.7 \\
& \textcolor{nvidiaGreen}{AF-Next-Instruct} & & \textbf{86.2}
\\ \cmidrule{1-4}

\multirow{2}{*}{\textbf{CMM Hallucination}}
& \textcolor{nvidiaGreen}{Audio Flamingo 3} & \multirow{2}{*}{ACC $\uparrow$} & 86.5 \\
& \textcolor{nvidiaGreen}{AF-Next-Instruct} & & \textbf{87.0} \\\cmidrule{1-4}

\multirow{2}{*}{\textbf{CompA-R-\textit{test}}}
& \textcolor{nvidiaGreen}{Audio Flamingo 3} &\multirow{2}{*}{ACC $\uparrow$}  & 98.0 \\
& \textcolor{nvidiaGreen}{AF-Next-Instruct} & & \textbf{98.7}   \\\cmidrule{1-4}

\multirow{2}{*}{\textbf{LibriSQA}}
& \textcolor{nvidiaGreen}{Audio Flamingo 3} & \multirow{2}{*}{GPT4o $\uparrow$} & 8.7 \\
& \textcolor{nvidiaGreen}{AF-Next-Instruct} & & \textbf{9.3}   \\ \cmidrule{1-4}
\multirow{2}{*}{\shortstack[l]{\textbf{NSynth} \\ \textit{Source | Instrument} }}
& \textcolor{nvidiaGreen}{Pengi} | \textcolor{qwenPurple}{Qwen-A} & \multirow{2}{*}{ACC ↑} & 62.0 | 78.8 \\
& \textcolor{nvidiaGreen}{AF-Next-Instruct} & & \textbf{66.7} | \textbf{81.7} \\ \cmidrule{1-4}

\multirow{2}{*}{\shortstack[l]{\textbf{Medley-Solos-DB} \\ \textit{Instrument} }}
& \textcolor{nvidiaGreen}{Audio Flamingo 2} & \multirow{2}{*}{ACC $\uparrow$} &  85.80 \\
& \textcolor{nvidiaGreen}{AF-Next-Instruct} & &  \textbf{92.13} \\ \cmidrule{1-4}

\multirow{2}{*}{\textbf{MuchoMusic}}
& \textcolor{nvidiaGreen}{Music Flamingo} & \multirow{2}{*}{ACC $\uparrow$} & 74.5 \\
& \textcolor{nvidiaGreen}{AF-Next-Instruct} & & \textbf{75.6}   \\

\cmidrule{1-4}

\multirow{2}{*}{\shortstack[l]{\textbf{SongCaps} \\ \textit{GPT5-Coverage \textbar{} GPT5-Correctness} }}
& \textcolor{nvidiaGreen}{Audio Flamingo 3} & \multirow{2}{*}{GPT5 $\uparrow$} &   6.7 \textbar{} 6.2 \\
& \textcolor{nvidiaGreen}{AF-Next-Instruct} & & \textbf{8.8} \textbar{} \textbf{8.9} \\ \cmidrule{1-4}
\multirow{3}{*}{\textbf{LongAudioBench}}
& \textcolor{closedGray}{Gemini-2.5-Pro} & \multirow{2}{*}{GPT4o $\uparrow$} & 60.4 \\
& \textcolor{nvidiaGreen}{Audio Flamingo 3} & & 68.6   \\ 
& \textcolor{nvidiaGreen}{AF-Next-Instruct} & & \textbf{73.9}   \\ 

\multirow{3}{*}{\textbf{+Speech}}
& \textcolor{closedGray}{Gemini-2.5-Pro}  & \multirow{2}{*}{GPT4o $\uparrow$} & 66.2 \\
& \textcolor{nvidiaGreen}{Audio Flamingo 3} & &  72.9  \\ 
& \textcolor{nvidiaGreen}{AF-Next-Instruct} & &  \textbf{81.2}  \\ 
\cmidrule{1-4}

\multirow{3}{*}{\shortstack[l]{\textbf{LibriSpeech (en)} \\ \textit{test-clean \textbar{} test-other}}}
& \textcolor{qwenPurple}{Phi-4-mm} | \textcolor{qwenPurple}{Qwen2.5-O} & \multirow{3}{*}{WER $\downarrow$} & 1.67 | 3.4 \\ 
& \textcolor{nvidiaGreen}{Audio Flamingo 3} & & 1.57 | 3.13 \\ 
& \textcolor{nvidiaGreen}{AF-Next-Instruct} & & \textbf{1.54} | \textbf{2.76} \\ \cmidrule{2-4}
\multirow{3}{*}{\textbf{SPGISpeech (en)}} 
& \textcolor{qwenPurple}{Qwen2-A-Inst} & \multirow{3}{*}{WER $\downarrow$} & 3.0 \\
& \textcolor{nvidiaGreen}{Audio Flamingo 3} & & \textbf{1.86} \\ 
& \textcolor{nvidiaGreen}{AF-Next-Instruct} & & 1.91 \\ \cmidrule{2-4}
\multirow{3}{*}{\textbf{TEDLIUM (en)}} 
& \textcolor{qwenPurple}{Phi-4-mm} & \multirow{3}{*}{WER $\downarrow$} & \textbf{2.9} \\
& \textcolor{nvidiaGreen}{Audio Flamingo 3} & & 3.5 \\ 
& \textcolor{nvidiaGreen}{AF-Next-Instruct} & & 3.3 \\ \cmidrule{2-4}
\multirow{3}{*}{\textbf{GigaSpeech (en)}} 
& \textcolor{qwenPurple}{Phi-4-mm} & \multirow{3}{*}{WER $\downarrow$} & 9.8 \\
& \textcolor{nvidiaGreen}{Audio Flamingo 3} & & 10.2 \\ 
& \textcolor{nvidiaGreen}{AF-Next-Instruct} & & \textbf{9.8} \\ \cmidrule{2-4}
\multirow{3}{*}{\textbf{Common Voice 15 (en)}} 
& \textcolor{qwenPurple}{Phi-4-mm} & \multirow{3}{*}{WER $\downarrow$} & 7.6 \\
& \textcolor{nvidiaGreen}{Audio Flamingo 3} & & 7.4 \\ 
& \textcolor{nvidiaGreen}{AF-Next-Instruct} & & \textbf{7.2} \\ \cmidrule{2-4}
\multirow{3}{*}{\textbf{VoxPopuli (en)}} 
& \textcolor{qwenPurple}{Phi-4-mm} & \multirow{3}{*}{WER $\downarrow$} & 5.9 \\
& \textcolor{nvidiaGreen}{Audio Flamingo 3} & & 5.6 \\
& \textcolor{nvidiaGreen}{AF-Next-Instruct} & & \textbf{5.4} \\ \bottomrule

\end{tabular}}
\vspace{-4mm}
\label{tab:main_results}
\end{table*}

\begin{table}[t]
\centering
\caption{\small Comparison of AF-Next-Instruct with open LALMs on VoiceBench and speech translation benchmarks.}
\resizebox{0.75\linewidth}{!}{%
\begin{tabular}{llcccc}
\toprule
\textbf{Task} & \textbf{Model} 
& \textbf{AdvBench} & \textbf{AlpacaEval} 
& \textbf{CommonEval} & \textbf{OpenBookQA} \\
\midrule

\multirow{3}{*}{\textbf{VoiceBench}}

& \textcolor{qwenPurple}{Qwen2.5-O} 
& \textbf{99.62} & 4.33 & 3.84 & 79.12  \\

& \textcolor{nvidiaGreen}{Audio Flamingo 3} 
& 98.26 & 4.19 & 3.40 & 66.81  \\
& \textcolor{nvidiaGreen}{AF-Next-Instruct} 
& 98.84 & \textbf{4.43} & \textbf{3.96} & \textbf{80.9}  \\

\midrule

\multicolumn{6}{c}{\textbf{CoVoST2 (Speech Translation, BLEU $\uparrow$)}} \\
\midrule

\textbf{Lang.} & \textbf{Model} 
& \textbf{ZH} & \textbf{JA} & \textbf{AR} & \textbf{DE}  \\
\midrule

\multirow{2}{*}{\textbf{EN $\rightarrow$ X}}

&  \textcolor{qwenPurple}{Phi-4-mm}
& 38.0 & \textbf{31.9} & 9.9 & \textbf{35.3} \\
& \textcolor{nvidiaGreen}{AF-Next-Instruct} 
& \textbf{38.2} & 29.6 & \textbf{21.9} & 31.4  \\
\midrule

\multirow{2}{*}{\textbf{X $\rightarrow$ EN}}

&  \textcolor{qwenPurple}{Phi-4-mm}
& 24.9 & \textbf{33.3} & 5.5 & \textbf{37.9} \\
& \textcolor{nvidiaGreen}{AF-Next-instruct} 
& \textbf{25.6} & 27.2 & \textbf{29.4} & 33.0  \\
\bottomrule
\end{tabular}}
\label{tab:voice_results_updated}
\end{table}

\noindent \textbf{Sequence Packing. }We employ a three-stage packing strategy to handle heterogeneous sequence lengths: (i) \textit{SP-Aware Sampling}, where the distributed sampler partitions data across data-parallel (DP) groups while ensuring all GPUs within an SP group receive identical sample indices. With SP degree $P$, the effective DP replica count reduces to        
$N_{\text{GPU}} / P$. Indices from each SP rank are interleaved so that every rank in a group loads the same batch at each step. A batch-level shuffle provides stochasticity without breaking this alignment; (ii) \textit{Padding and Truncation}, where the data collator pads all sequences in a batch to the shorter of the longest sequence and the maximum context length, constructs a binary attention mask over non-padding positions, and pads labels with an ignore index; and (iii) \textit{Audio Token Expansion}, where during audio encoding stage, each audio placeholder token is replaced by a variable number of audio embedding tokens determined by the clip's duration-based embedding mask.

\noindent \textbf{Hybrid Sequence Parallelism. }We distribute attention across $P$ GPUs using Unified Sequence Parallelism (USP), decomposed into a Ulysses degree $P_U$ (all-to-all based) and a Ring degree $P_R$ (point-to-point based), with $P = P_U \times P_R$. The system constructs separate NCCL process groups for each: a Ulysses group, a Ring group, and a Data-Parallel group.  Ulysses attention~\citep{jacobs2023deepspeed} redistributes the sequence and head dimensions across GPUs via all-to-all collectives, giving each GPU the full sequence but only a fraction of the attention heads (as shown in ~\cref{fig:architecture}) --efficient within high-bandwidth interconnects but costly across nodes. Ring attention~\citep{liu2023ring} instead circulates KV blocks around a ring topology via point-to-point transfers, scaling across nodes but introducing sequential latency proportional to the ring size. Hybrid SP composes both: Ulysses operates within nodes where all-to-all bandwidth is abundant, while Ring spans across nodes, keeping communication efficient at both levels~\citep{fang2024usp}.          

%% file: sections/05.experiments.tex
\section{Experiments}
\label{sec:experiments}

{\noindent \textbf{Experimental Setup.}} We perform pre-training, mid-training, post-training, and CoT-training of AF-Next on 128 NVIDIA H100 GPUs. Further details on batch size, learning rates, and optimizers for each stage of training are in Appendix~\ref{sec.afnext_training_details}. To evaluate AF-Next Captioner, we use the model to generate a caption for the audio and prompt GPT-5.2 in text-only mode with the caption and the associated question.

\noindent \textbf{Baselines.} We evaluate all 3 of our model variants against recent SOTA LALMs, including GAMA~\citep{ghosh2024gama}, Audio Flamingo~\citep{kong2024audio}, Audio Flamingo 2, Audio Flamingo 3, Qwen-A(udio)~\citep{chu2023qwenaudio}, Qwen2-A(udio)~\citep{chu2024qwenaudio2}, Qwen2-A(udio)-(Inst)ruct, Qwen2.5-O(mni)~\citep{xu2025qwen2}, Qwen3-O(mni)~\citep{xu2025qwen3}, R1-AQA~\cite{li2025reinforcement}, Pengi~\citep{deshmukh2023pengi}, Phi-4-mm~\citep{abouelenin2025phi}, Baichun Audio~\citep{li2025baichuan}, Step-Audio-Chat~\citep{huang2025step}, LTU~\citep{gong2023listen}, LTU-AS~\citep{gong2023ltu-as}, SALMONN~\citep{tang2023salmonn}, AudioGPT~\citep{huang2023audiogpt}, and Gemini (2.0 Flash, 1.5 Pro, 2.5 Flash and 2.5 Pro)~\citep{team2023gemini} (note we do not evaluate Gemini on ASR benchmarks due to low rate limits), as well as GPT-4o-audio~\citep{hurst2024gpt}. For LongAudioBench, for models that do not support longer audio, we follow the cascaded approach for evaluation proposed by~\citet{ghosh2025audio}. We run all the mentioned baselines, and we report reproduced scores.

\noindent \textbf{Evaluation Datasets.} We evaluate our AF-Next series of models on a variety of tasks and benchmarks, including \textit{audio classification} (NSynth (Source and Instrument)~\citep{engel2017neural}, NonSpeech7k~\citep{rashid2023nonspeech7k}, LibriSQA~\citep{zhao2023librisqa}), \textit{reasoning-focused audio QA} (MMAU~\citep{sakshi2024mmau} (v05.15.25), MMAU-Pro~\citep{kumar2025mmauprochallengingcomprehensivebenchmark},  MuchoMusic (perceptual version)~\citep{zang2025you,weck2024muchomusic}, MMAR~\citep{ma2025mmarchallengingbenchmarkdeep}, MMSU~\citep{wang2025mmsu}, CompA-R-test~\citep{ghoshcompa}), \textit{multimodal hallucination detection} (CMM~\citep{leng2024curse}), \textit{ASR} (Librispeech (clean and other)~\citep{panayotov2015librispeech}, SPGISpeech~\citep{o2021spgispeech}, TEDLIUM~\citep{rousseau2012ted,hernandez2018ted}, and Voxpopuli~\citep{wang2021voxpopuli}), LongAudioBench~\citep{ghosh2025audio} and SongCaps~\citep{ghosh2025music}. To calculate accuracy, we use either exact string matching with the ground truth or CLAP-based retrieval following~\citep{deshmukh2023pengi}, implemented with open-source AF-CLAP~\citep{ghosh2025audio}. For MCQ, AF-Next typically outputs only the selected option. In cases where the model provides more verbose or open-ended responses (e.g., AF-Next-Think), we apply multiple regex patterns to extract the chosen option. Although AF-Next supports a broader range of capabilities, including multi-talker ASR, speaker diarization, timestamped captioning, and voice-to-voice interaction, etc, we restrict this submission to the most widely used benchmarks and leave evaluation on these additional tasks to future work.

%% file: sections/06.results.tex
\section{Results}
\label{sec:results}

In \Cref{tab:main_results}, we present a comprehensive evaluation of Audio Flamingo Next across a diverse suite of audio understanding, reasoning, and speech recognition benchmarks. AF-Next-Instruct establishes itself as the strongest fully open-source LALM, substantially outperforming prior open models and remaining highly competitive with, while often surpassing, state-of-the-art open-weight and closed-source models on the majority of tasks. Furthermore, our AF-Next-Think and AF-Next-Captioner variants yield consistent additional gains, pushing performance even further. We present qualitative examples on our project website.
\vspace{0.5mm}

\noindent \textbf{Audio Understanding and Reasoning.} On MMAU-v05.15.25, AF-Next-Instruct achieves an average accuracy of 74.20, surpassing Audio Flamingo 3 (72.42). AF-Next-Think further improves this to 75.01, and incorporating the captioner pipeline AF-Next-Captioner yields the best result of 75.76, with gains across all three subcategories: sound (79.87), music (75.3), and speech (72.13). A similar trend holds on MMAR, where AF-Next-Instruct (59.7) already outperforms AF3 (58.5), and our AF-Next-Captioner variant pushes accuracy to 63.0 -- a 4.5-point absolute improvement over AF3. On MMSU, while the closed-source Gemini-2.5-Flash leads at 66.1, AF-Next narrows the gap substantially: our AF-Next-Captioner variant reaches 63.3, compared to 59.4 for the instruct variant. On the more challenging MMAU-Pro benchmark, AF-Next-Instruct (56.9) surpasses the closed-source Gemini-2.5-Pro (57.4), and AF-Next-Think extends this lead to 58.7. These results demonstrate that test-time compute strategies provide complementary benefits: CoT reasoning helps on tasks requiring multi-step inference, while captioner augmentation is particularly effective when richer acoustic descriptions can ground the model's reasoning.
\vspace{0.5mm}

\noindent \textbf{Audio Captioning, Entailment, and Classification.} AF-Next-Instruct improves audio captioning quality on both Clotho-v2 (CIDEr: 0.52 vs.\ 0.50) and AudioCaps (0.74 vs.\ 0.70) over AF3. On audio entailment, it achieves 94.2 on Clotho and 96.0 on AudioCaps, improving upon AF3's already strong results of 93.3 and 95.0, respectively. For sound event classification on NonSpeech7k, AF-Next reaches 86.2 accuracy (vs.\ 85.7 for AF3), and on the CMM Hallucination benchmark it scores 87.0 (vs.\ 86.5), indicating improved robustness to hallucinated audio content. On CompA-R, AF-Next achieves 98.7 accuracy, and on LibriSQA it reaches a GPT4o score of 9.3, both improvements over AF3.
\vspace{0.5mm}

\noindent \textbf{Music Understanding.} AF-Next demonstrates particularly strong gains on music benchmarks. On NSynth, it achieves 66.7 accuracy for source classification and 81.7 for instrument classification, outperforming the prior best open-source (Pengi, 62.0) and open-weight (Qwen-Audio, 78.8) models by substantial margins. On Medley-Solos-DB instrument recognition, AF-Next reaches 92.13, a notable improvement over Audio Flamingo 2's 85.80. On MuchoMusic, it scores 75.6 compared to Music Flamingo's 74.5. For music captioning on SongCaps, AF-Next achieves GPT5 coverage and correctness scores of 8.8 and 8.9, respectively, representing large improvements over AF3's 6.7 and 6.2.
\vspace{0.5mm}

\noindent \textbf{Long Audio Understanding.} On LongAudioBench, AF-Next-Instruct outperforms both AF3 (68.6) and the closed-source Gemini 2.5 Pro (60.4) by a wide margin, achieving 73.9. The gap is even more pronounced on the speech-inclusive variant (+Speech), where AF-Next reaches 81.2 compared to AF3's 72.9 and Gemini 2.5 Pro's 66.2. These results highlight AF-Next's strength in long-context audio and speech reasoning.
\vspace{0.5mm}

\noindent \textbf{Automatic Speech Recognition.} AF-Next-Instruct achieves competitive or state-of-the-art ASR performance across multiple English benchmarks. On LibriSpeech, it sets new lows among LALMs with a WER of 1.54 on test-clean and 2.76 on test-other, improving over both AF3 and open-weight models such as Phi-4-mm and Qwen2.5-Omni. It also achieves the best WER on Common Voice 15 (7.2), GigaSpeech (9.8), and VoxPopuli (5.4), while remaining competitive on SPGISpeech (1.91 vs.\ AF3's 1.86) and TEDLIUM (3.3 vs.\ Phi-4-mm's 2.9).

\noindent \textbf{Voice Understanding and Speech Translation.} We further evaluate AF-Next-Instruct on VoiceBench and speech translation tasks in \Cref{tab:voice_results_updated}. On VoiceBench, AF-Next-Instruct achieves the highest scores on AlpacaEval (4.43), CommonEval (3.96), and OpenBookQA (80.9), outperforming both the open-weight Qwen2.5-Omni and the open-source AF3 across these subtasks. Notably, on OpenBookQA, AF-Next surpasses AF3 by over 14 points and edges out Qwen2.5-Omni (79.12), while maintaining a strong AdvBench safety score of 98.84. On CoVoST2 speech translation, AF-Next demonstrates competitive multilingual capabilities against Phi-4-mm. For EN$\rightarrow$X translation, AF-Next achieves the best BLEU scores on Chinese (38.2) and Arabic (21.9) --- the latter representing a substantial 12-point improvement over Phi-4-mm (9.9) --- while remaining competitive on Japanese and German. A similar pattern emerges for X$\rightarrow$EN translation, where AF-Next leads on Chinese (25.6) and Arabic (29.4), with the Arabic result again showing a dramatic improvement over Phi-4-mm (5.5). These results suggest that AF-Next's multilingual speech understanding is particularly strong for underrepresented language pairs such as Arabic, while maintaining competitive performance on higher-resource languages.

%% file: sections/07.conclusion.tex
\section{Conclusion}
\label{sec:conclusion}
\vspace{-1mm}
In this paper, we present Audio Flamingo Next (AF-Next), the most capable model in the Audio Flamingo series to date. Beyond achieving SOTA performance on a wide range of contemporary audio understanding benchmarks, AF-Next demonstrates substantially stronger robustness to real-world use cases and supports a broad set of capabilities, including understanding long-form audio of up to 30 minutes, multi-turn chat, timestamped captioning, and multilingual ASR. We open-source our training code, model checkpoints, and core techniques to support future research in open audio-language modeling. In addition, we introduce Temporal Audio Chain-of-Thought, a new reasoning paradigm for long-audio question answering that explicitly grounds intermediate evidence in time, enabling more faithful and robust reasoning.

%% file: sections/08.limitations.tex
\section*{Limitations}
AF-Next has several important limitations. First,
although we substantially scale training data beyond prior open audio-language models, internet-scale audio remains noisy and unevenly distributed across domains, languages, and acoustic conditions. In particular, low-resource languages, rare sound
events, and specialized real-world domains are still underrepresented. Future work should focus on improving the diversity, balance, and coverage of open audio datasets.

Second, while AF-Next improves long-audio understanding and supports audio up to 30 minutes,
robust reasoning over long contexts remains challenging when evidence is temporally distant, sparse,
or distributed across multiple segments. Although
Temporal Audio Chain-of-Thought improves temporal grounding, stronger long-context memory,
retrieval, and evidence aggregation remain important directions for future work.

Third, our evaluation focuses on the most established benchmarks, and therefore does not yet fully
cover several capabilities supported by AF-Next, including multi-talker ASR, speaker diarization,
timestamped captioning, and voice-to-voice interaction. Building broader evaluation protocols for
these capabilities is an important next step.

%% file: sections/__appendix.tex
\section{Appendix}
\begin{figure*}[h]
  \centering
  \includegraphics[width=\textwidth]{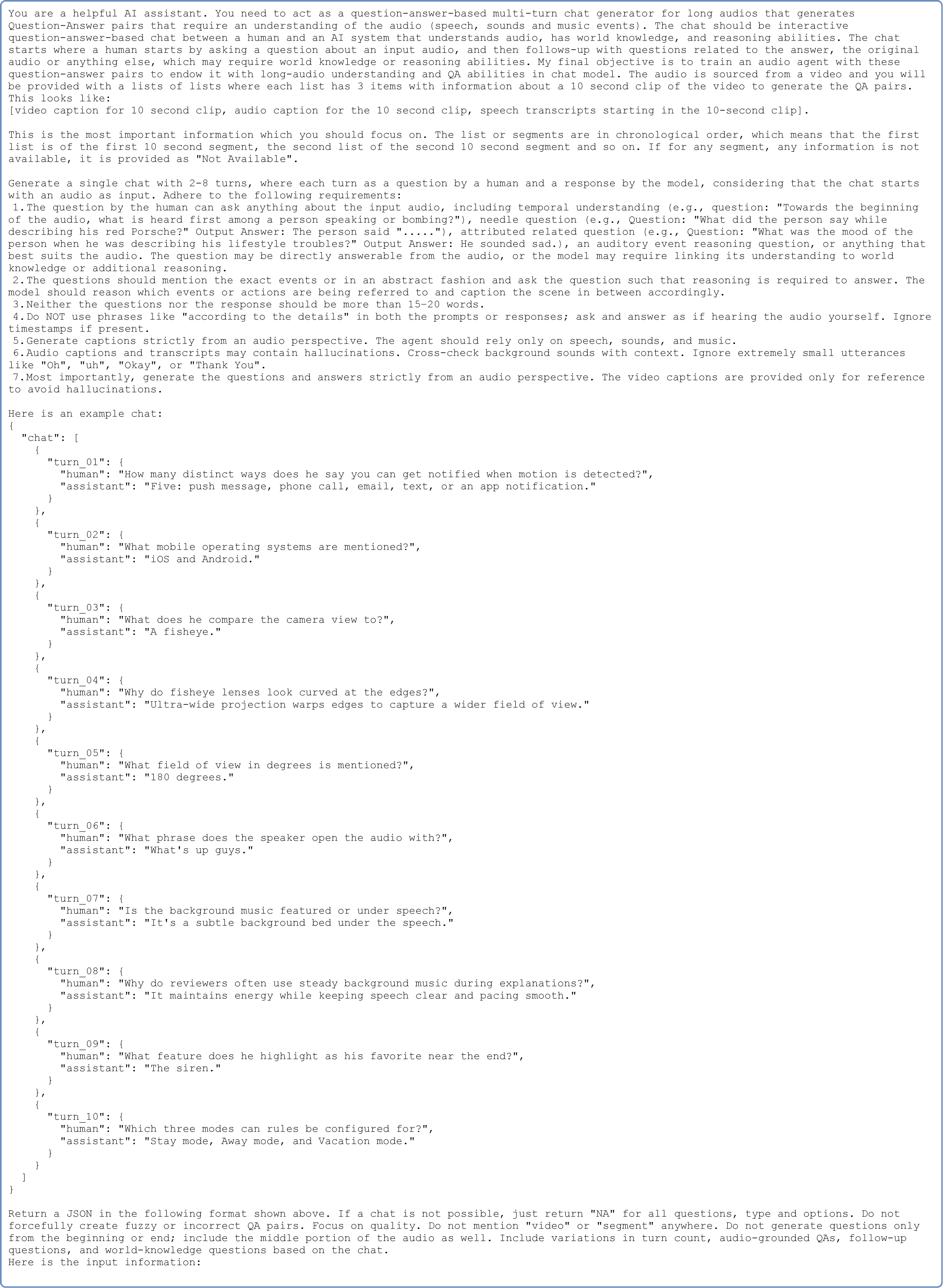}
  \caption{Prompt used for generating multi-turn chat QA pairs from long audio.}
  \label{fig:prompt_chat_audio_only}
  \vspace{-0.2cm}
\end{figure*}

\begin{figure*}[h]
  \centering
  \includegraphics[width=\textwidth]{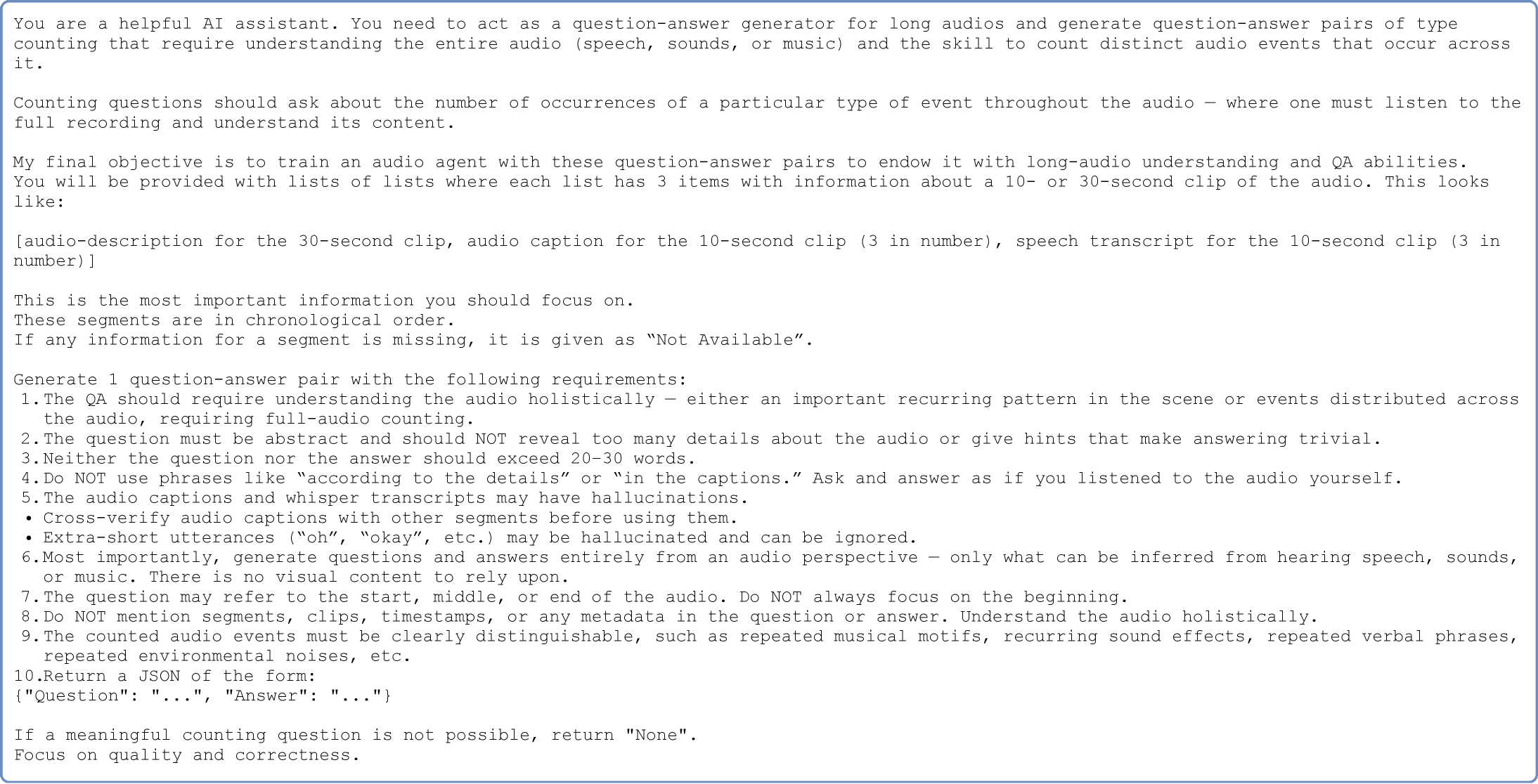}
  \caption{Prompt used for generating counting QA pairs from long audio.}
  \label{fig:prompt_counting}
  \vspace{-0.2cm}
\end{figure*}

\begin{figure*}[h]
  \centering
  \includegraphics[width=\textwidth]{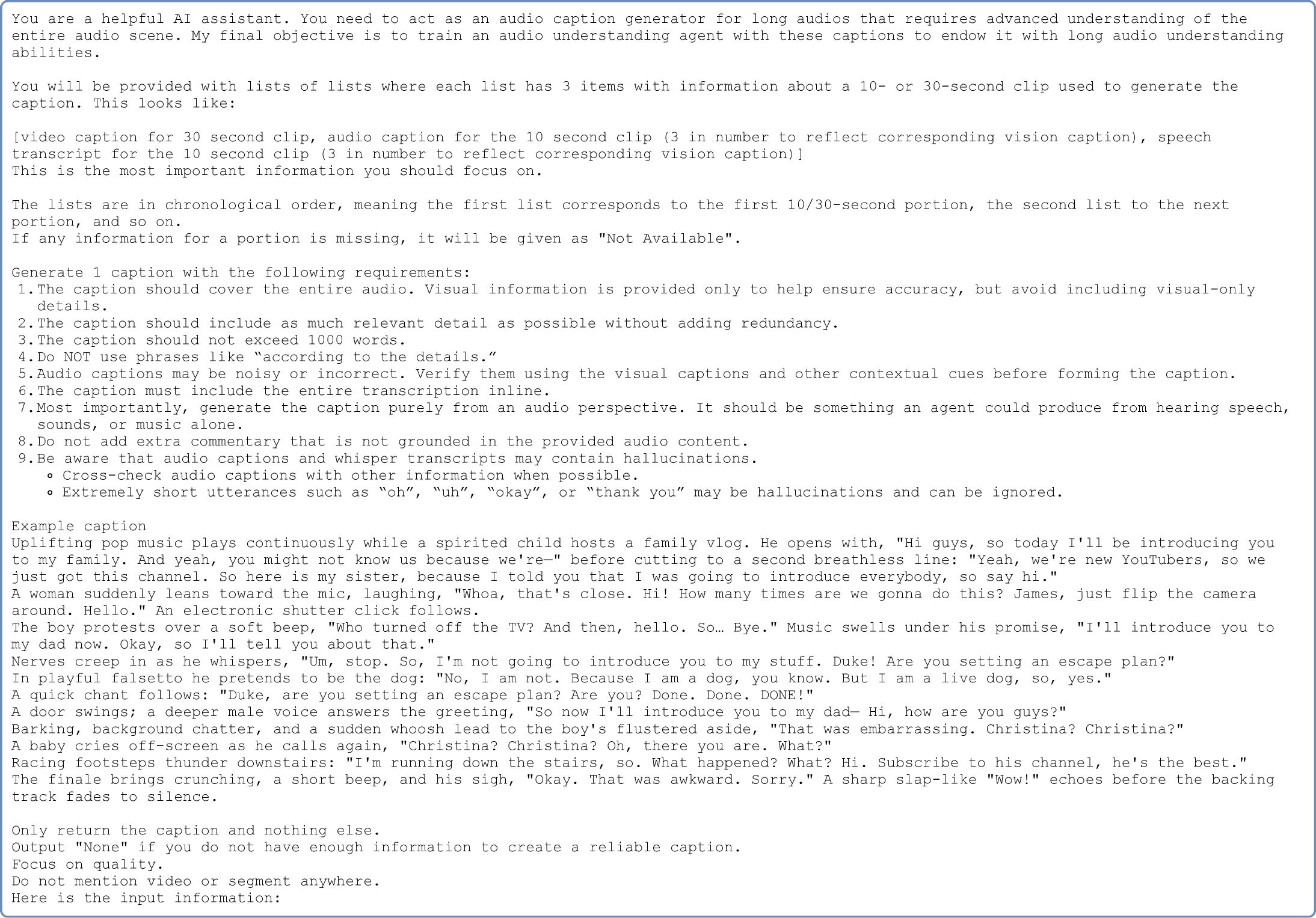}
  \caption{Prompt used for generating detailed audio captions from long audio.}
  \label{fig:prompt_detailed_caption_audio_only}
  \vspace{-0.2cm}
\end{figure*}

\begin{figure*}[h]
  \centering
  \includegraphics[width=\textwidth]{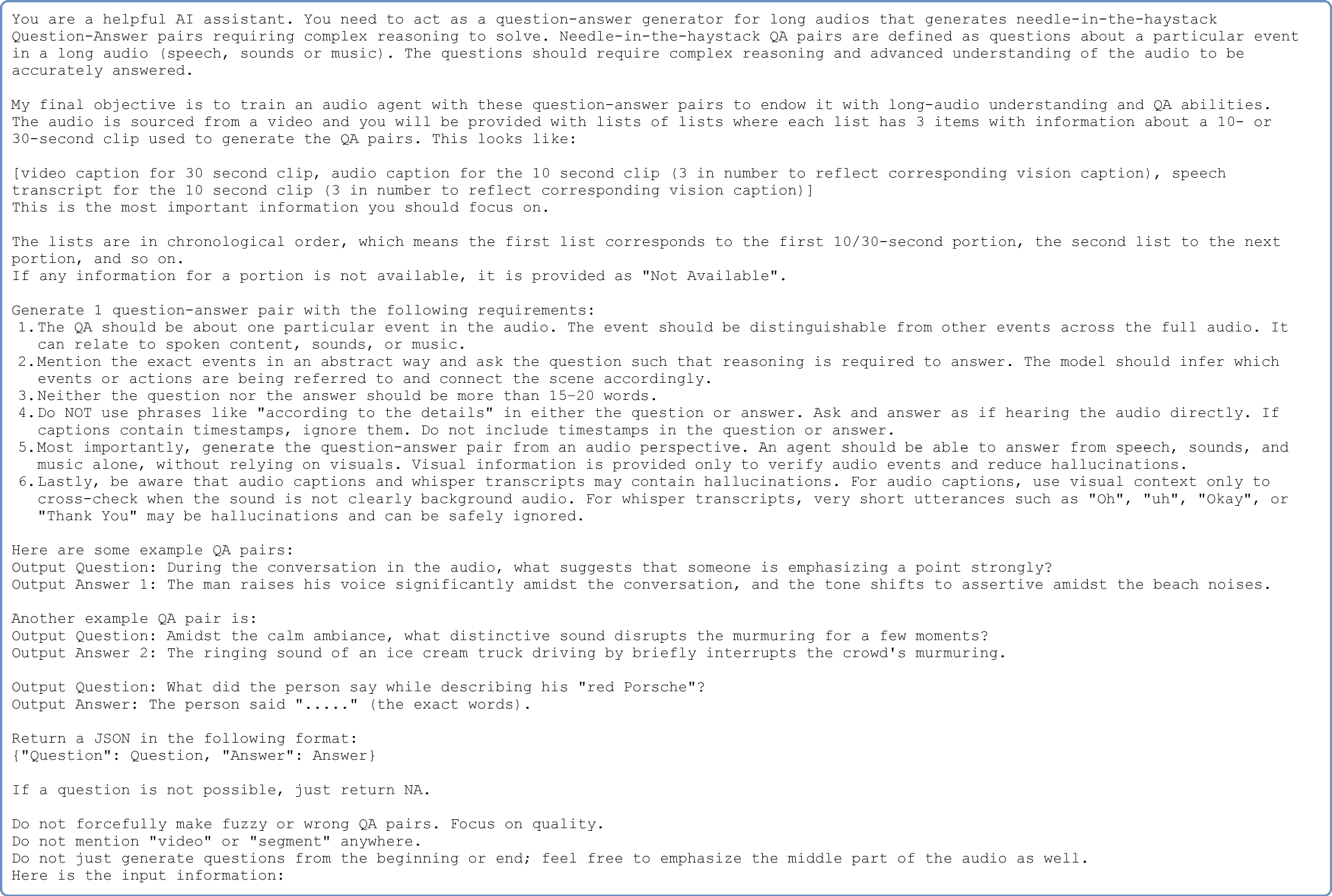}
  \caption{Prompt used for generating needle-in-the-haystack QA pairs from long audio.}
  \label{fig:prompt_needle_audio_only}
  \vspace{-0.2cm}
\end{figure*}

\begin{figure*}[h]
  \centering
  \includegraphics[width=\textwidth]{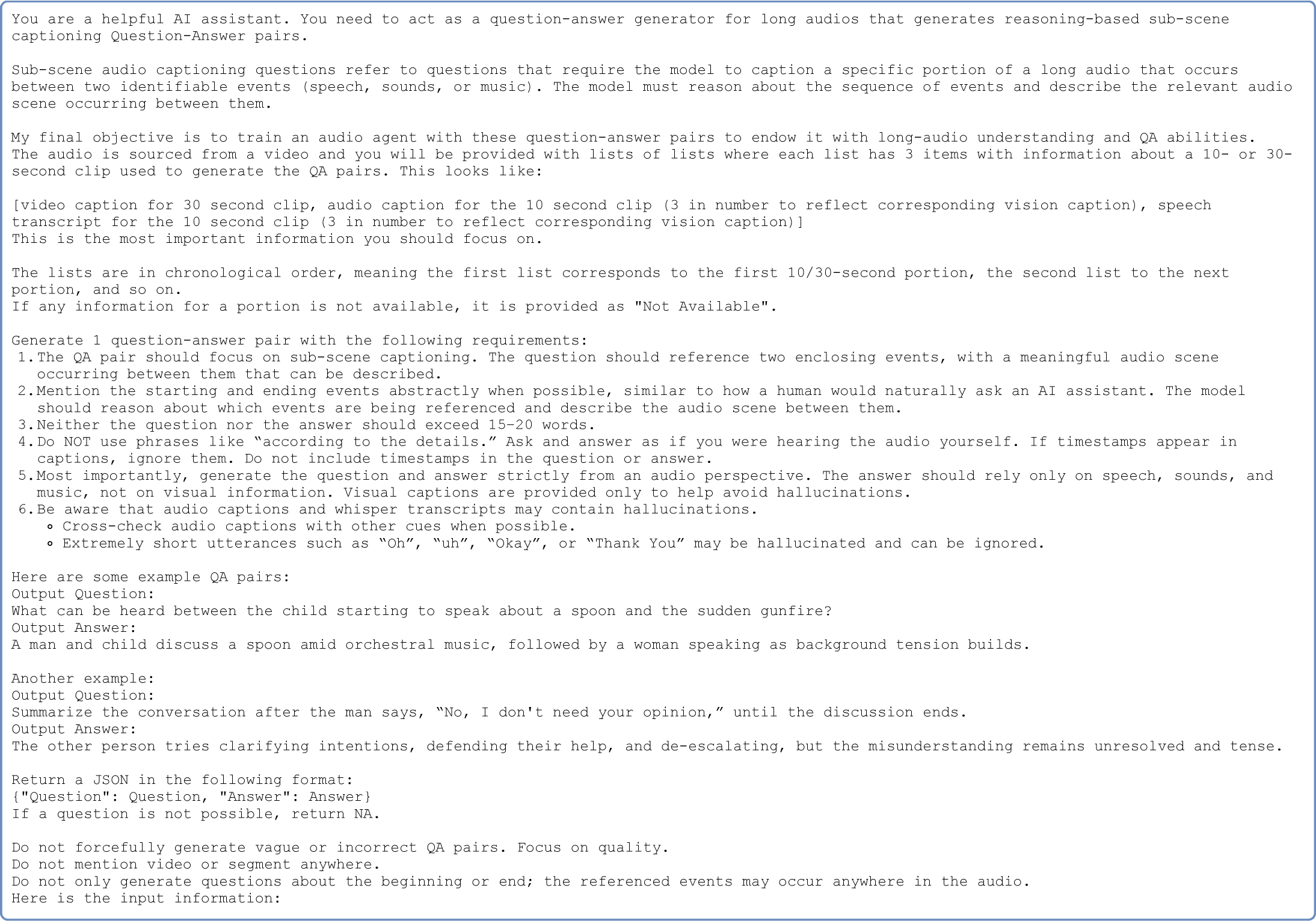}
  \caption{Prompt used for generating subscene captioning QA pairs from long audio.}
  \label{fig:prompt_subscene_audio_only}
  \vspace{-0.2cm}
\end{figure*}

\begin{figure*}[h]
  \centering
  \includegraphics[width=\textwidth]{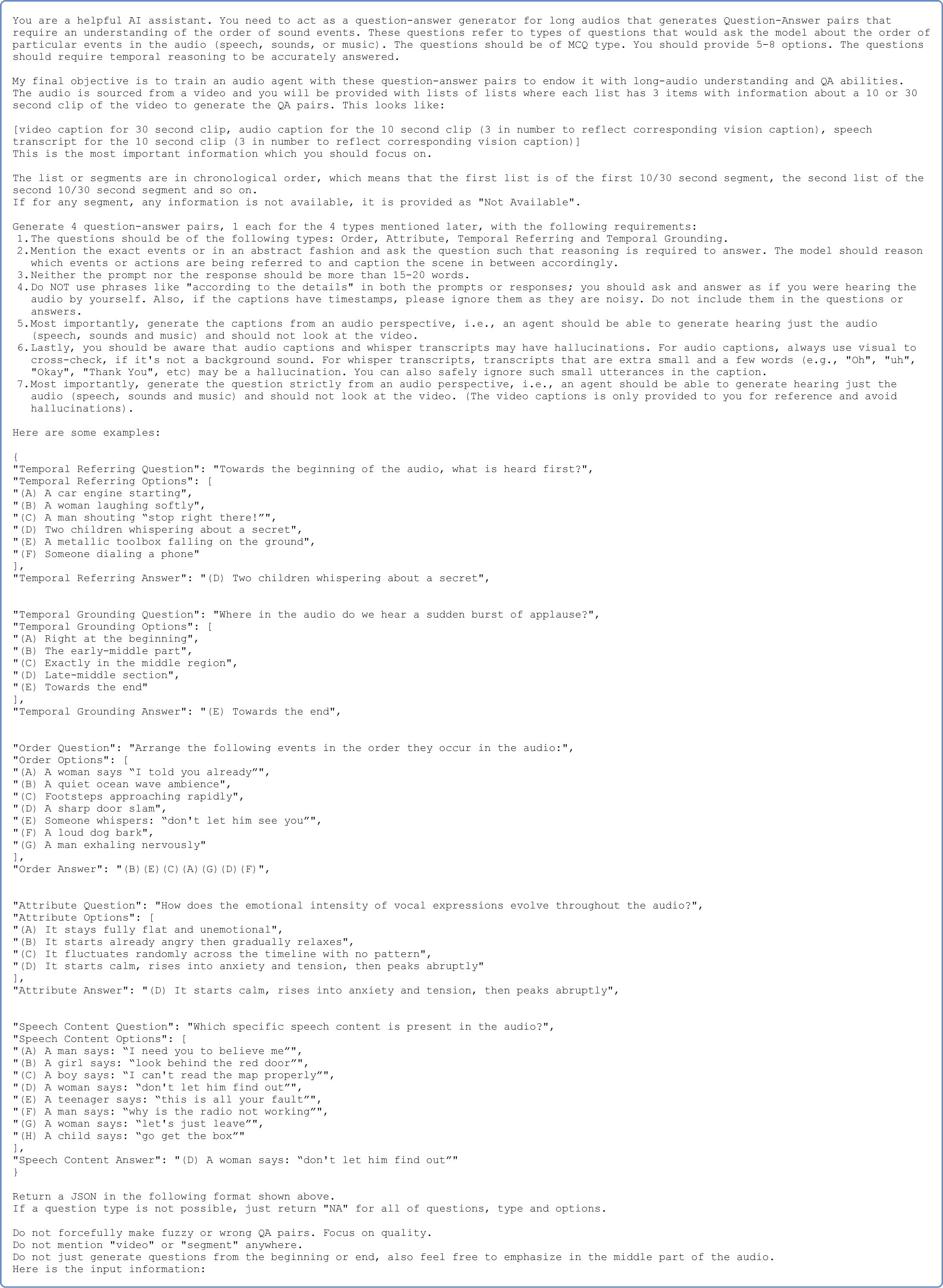}
  \caption{Prompt used for generating temporal understanding QA pairs from long audio.}
  \label{fig:prompt_temporal_audio_only}
  \vspace{-0.2cm}
\end{figure*}

\begin{figure*}[h]
  \centering
  \includegraphics[width=\textwidth]{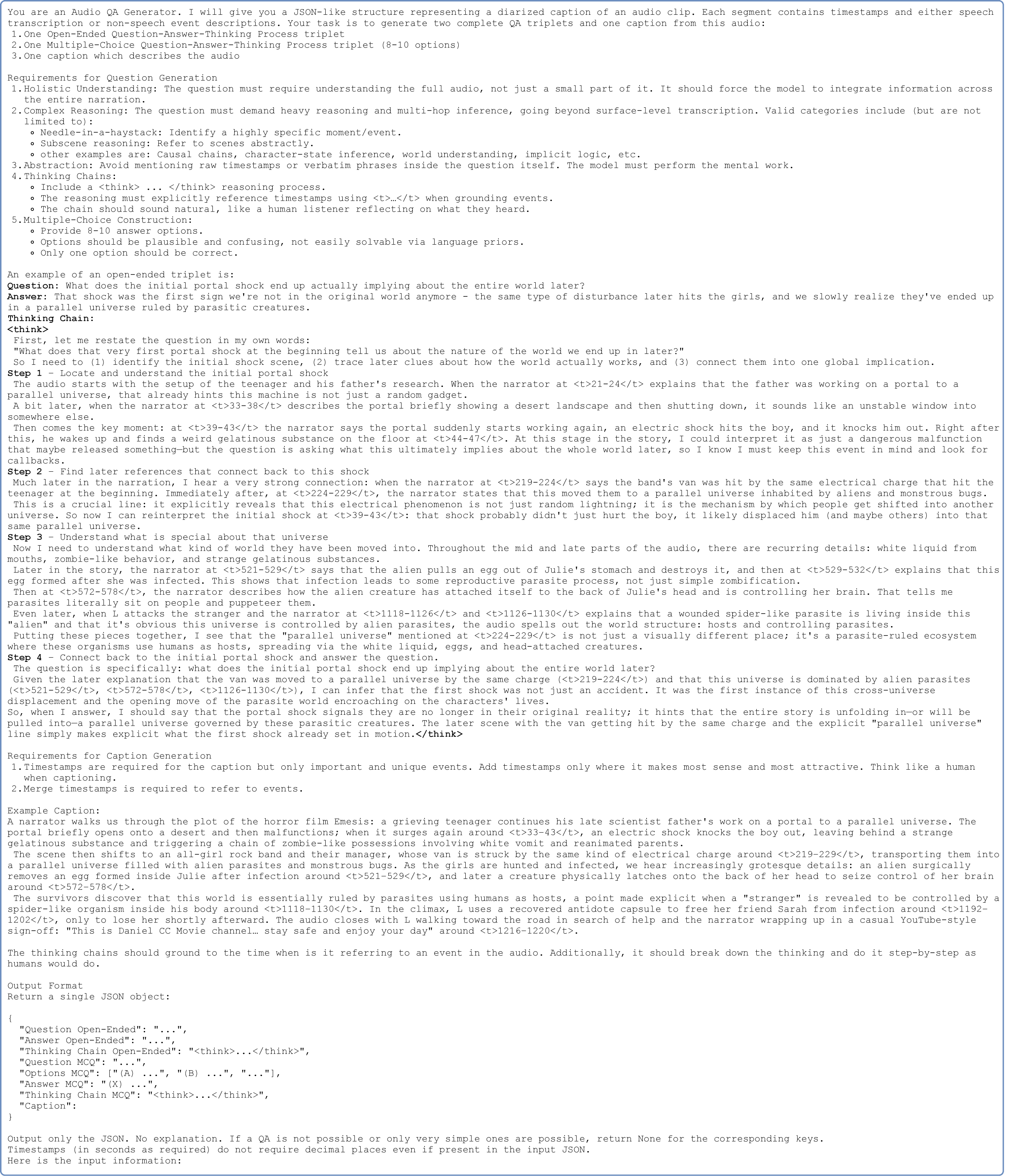}
  \caption{Prompt used for generating time-grounded chain-of-thought reasoning and timestamped caption data.}
  \label{fig:prompt_thinking}
  \vspace{-0.2cm}
\end{figure*}

\begin{figure*}[h]
  \centering
  \includegraphics[width=\textwidth]{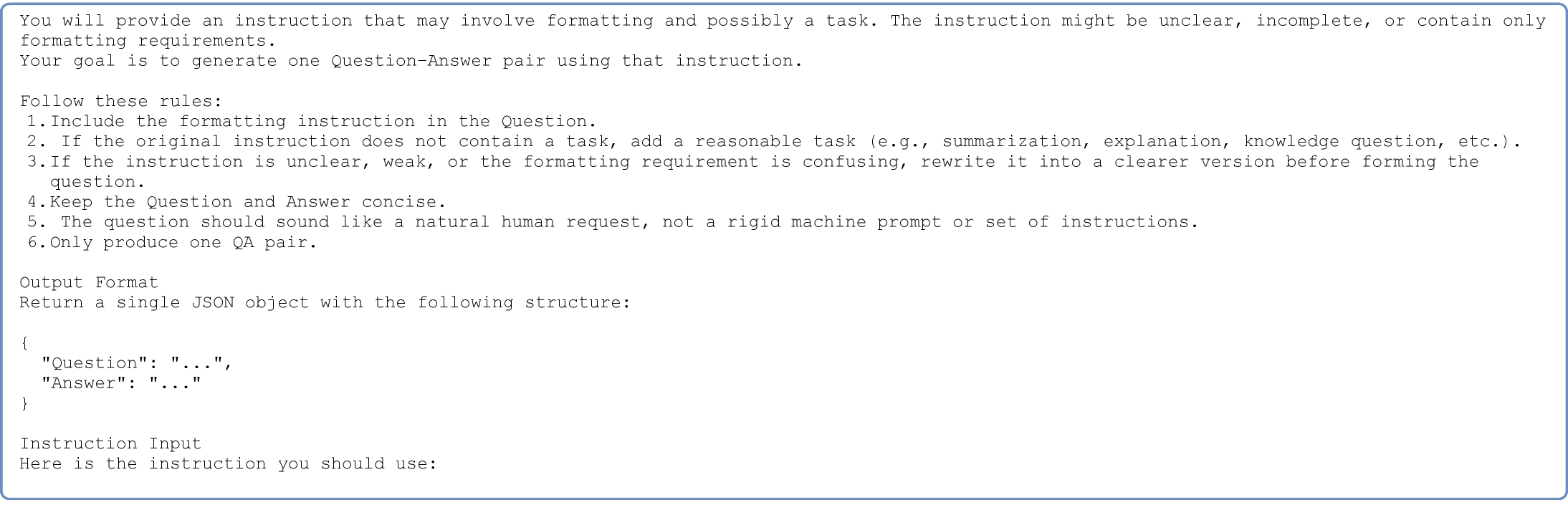}
  \caption{Prompt used for generating instruction-following QA pairs.}
  \label{fig:prompt_instruction_following}
  \vspace{-0.2cm}
\end{figure*}

In this appendix, we provide additional details on Potential Risks (Appendix~\ref{app:potential_risks}), data anonymization and privacy safeguards (Appendix~\ref{app:pii}), training configurations (Appendix~\ref{sec.afnext_training_details}), dataset composition (Appendix~\ref{app:dataset_details}), descriptive statistics (Appendix~\ref{app:descriptive_stats}), and data generation prompts (Figures~\ref{fig:prompt_chat_audio_only}--\ref{fig:prompt_instruction_following}).

\section{Potential Risks}
\label{app:potential_risks}

AF-Next inherits the general risks associated with large language models, including the potential to generate toxic, biased, or hallucinated outputs. Since the model is trained on internet-scale audio and text data, it may reflect biases present in the training distribution, such as varying recognition accuracy across accents and dialects. Additionally, as with any generative model, AF-Next may occasionally produce plausible but factually incorrect responses, particularly for long-audio inputs where evidence must be aggregated across extended temporal spans. To mitigate these risks, we include 386K safety and instruction-following fine-tuning samples that teach the model appropriate refusal and abstention behavior.

\section{Data Anonymization and Privacy Safeguards}
\label{app:pii}

Our training data is derived from publicly available academic datasets and openly accessible internet audio. We do not intentionally collect data that names or uniquely identifies individuals. As additional safeguards, we strip metadata that could link audio samples to real-world speaker identities during data curation, and we include 386K safety fine-tuning samples that train the model to refuse requests for extracting personally identifiable information (Section~\ref{subsubsec:data_curation}, Figure~\ref{fig:datasets_example}).

\section{AF-Next Training Details}
\label{sec.afnext_training_details}
In this section, we present the training settings of our model across all stages, each with specific configurations. Details are in \cref{tab:blend_weights}.

\begin{table}[!ht]
\centering
\resizebox{\linewidth}{!}{
\begin{tabular}{lcccc}
\toprule
\textbf{Settings} & \textbf{Pre-training} & \textbf{Mid-Training} & \textbf{Post-Training} & \textbf{CoT-Training} \\
\midrule
global batch size & 128 & 128 & 64 & 64 \\
learning rate & 1e-3 & 1e-5 & 1e-6 & 2e-5\\
learning schedule & \multicolumn{4}{c}{Cosine decay} \\
warm up ratio & \multicolumn{4}{c}{0.03}\\
weight decay & \multicolumn{4}{c}{0.0}\\
epoch & 1 & 1 & 1 & 2\\
bf16 & \checkmark & \checkmark & \checkmark & \checkmark \\
grad accumulate & \multicolumn{4}{c}{8} \\
Parallelism  & Zero-3 & Zero-3 + SP & DP & Zero-3 + SP \\
GPUs & \multicolumn{4}{c}{128$\times$H100} \\
\bottomrule
\end{tabular}}
\caption{\small Training settings across stages.}
\label{tab:blend_weights}

\end{table}

\section{AF-Next Training Dataset Details}
\label{app:dataset_details}

Table~\ref{tab:dataset-details} summarizes all datasets used to train AF-Next, including total hours, number of audio-QA pairs, and the number of epochs (passes over the dataset) used at each training stage. 
We convert all foundational datasets (captioning, classification, etc.) into QA formats, using the same set of prompts for each task as mentioned in ~\citep{ghosh2025audio, goel2025audio}.

\begin{table*}[h]
    \centering
    \caption{\small List of pre-training and fine-tuning datasets together with their training composition.}
    \resizebox{\textwidth}{!}{
    \begin{tabular}{lcccccc}
    \toprule
Dataset & Hours  & Num. Pairs & Pre-training & Mid-Training & Post-Training & CoT-Training \\ \midrule
        Long Audio Captioning (Ours) & 27k hrs & 290K   & - & 2.0 & 1.0 & - \\
        Long Temporal QA (Ours) & 39k hrs & 317K   & - & 2.0 & 1.0 & -  \\
      Long Needle QA (Ours)& 34k hrs & 281K   & - & 2.0 & 1.0 & -  \\
    Long Subscene QA (Ours)& 28k hrs & 256K   & - & 2.0 & 1.0 & - \\
      Long Counting QA (Ours)& 26k hrs & 170K   & - & 2.0 & 1.0 & -  \\
      Safety QA (Ours) & 536 hrs & 386K   & - & 2.0 & 1.0 & - \\
      Multilingual QA (Ours)& 82 hrs & 45K   & - & 2.0 & 1.0 & -  \\
      Speaker Analysis (Ours)& 56 hrs & 11K   & - & 2.0 & 1.0 & -  \\
      Audio Comparison (Ours)& 115 hrs & 30K   & - & 2.0 & 1.0 & -   \\
      Acoustic Quality (Ours) & 4 hrs &  1500  & - & 2.0 & 1.0 & - \\
        AF3-training mix~\citep{goel2025audio} & 308k hrs & 48.6M & 1.0 & 1.0 & - & - \\
        AudioSkills-XL ~\citep{goel2025audio} & - & 9700K & - & 2.0 & 1.0 & - \\
        MF-Skills~\citep{ghosh2025music} & - & 3M & - & 2.0 & 1.0 & - \\
        MusicBench~\citep{melechovsky2023mustango} & 115.5 hrs & 686k & 1.0 & 1.0& - & -\\
        Mu-LLAMA~\citep{liu2024music} & 62.9 hrs & 70k & 1.0 & 1.0& - & -\\
        MusicAVQA\textsubscript{audio-only}~\citep{li2022learning} & 77.1 hrs & 5.7K & 1.0 & 1.0& - & -\\
        MusicQA~\citep{ouyang2025mqad} & 62.9 hrs & 70K & 1.0 & 1.0 & - & - \\
        LP-MusicCaps\textsubscript{MSD}~\citep{doh2023lp} & 5805.7 hrs & 1331.8K & 1.0 & 1.0& - & -\\
        LP-MusicCaps\textsubscript{MTT}~\citep{doh2023lp} & 126.4 hrs & 46.9K & 1.0 & 1.0& - & -\\
        LP-MusicCaps\textsubscript{MC}~\citep{doh2023lp} & 7.4 hrs & 7.9K & 1.0 & 1.0& -& -\\
        MusicCaps~\citep{agostinelli2023musiclmgeneratingmusictext} & 7.4 hrs & 2.6K & 1.0 & 1.0& -& -\\
        NSynth~\citep{engel2017neural} & 321.3 hrs & 289.2K & 1.0 & 1.0& - & -\\
        MusDB-HQ~\citep{rafii2017musdb18} & 29.1 hrs & 10.2K & 1.0 & 1.0& - & -\\
        FMA~\citep{defferrard2016fma} & 860.7 hrs & 104.2K & 1.0 & 1.0& - & -\\
    
        Music4All Captions~\citep{ghosh2025music}& 910.5 hrs & 109k  & 1.0 & 1.0&- & -\\
        Music4All QA~\citep{ghosh2025music}& 1505.7 hrs & 180k & 1.0 & 1.0& 1.0 & - \\
        MSD Captions~\citep{ghosh2025music}& 15449.9 hrs & 1.4M & 1.0 & 1.0& - & -\\
        MSD QA~\citep{ghosh2025music}& 20906.2 hrs & 935k & 1.0 & 1.0& 1.0 & - \\

        CHIME~\citep{foster2015chime} & 342 hrs & 30k & $1.0$ & 1.0 &- & -\\
        ALI Meeting~\citep{yu2022m2meticassp2022multichannel} & 118.75 hrs & 387k & $1.0$ & 1.0 &- & -\\
        EMILIA~\citep{he2024emiliaextensivemultilingualdiverse} & 5000 hours & 1.7M & $1.0$ & 1.0 &- & -\\
        MUST~\citep{qin2025mustdatasetunifiedframework} & 500 hrs & 245k & $1.0$ & 1.0 &- & - \\
        CoVoST~\citep{wang2020covost2massivelymultilingual} & 2880 hrs & 5M & $1.0$ & 1.0 &-& -\\
        Aidatatang~\citep{aidatatang200zh} & 139.39 hrs  & 165k & $1.0$ & 1.0 &- & -\\
        Aishell~\citep{aishell_2017} & 150.85 hrs & 120k & $1.0$ & 1.0 &- & -\\
        Multi-talker Switchboard~\citep{godfrey1992switchboard} & 109.9 hrs & 76.6K & $1.0$ & 1.0 & -& -\\
        AF-Think-Time (Ours) & 3954 hrs & 43K & - &  - &  - & 2.0 \\
       
    \bottomrule
    \end{tabular}}
    
    \label{tab:dataset-details}
\end{table*}

\section{Descriptive Statistics}
\label{app:descriptive_stats}

All results reported in Table~\ref{tab:main_results} are averaged over 3 independent runs. For accuracy-based metrics, we report mean accuracy. For ASR benchmarks, we report mean Word Error Rate (WER).

\section{AI Assistants Usage}
\label{app:ai_usage}
AI assistants were used only for grammar correction and language 
polishing during manuscript preparation. All scientific content, experimental 
design, methodology, analysis, and conclusions presented in this work are 
entirely the authors' own.

\begin{figure*}[h]
  \centering
  \includegraphics[width=\textwidth]{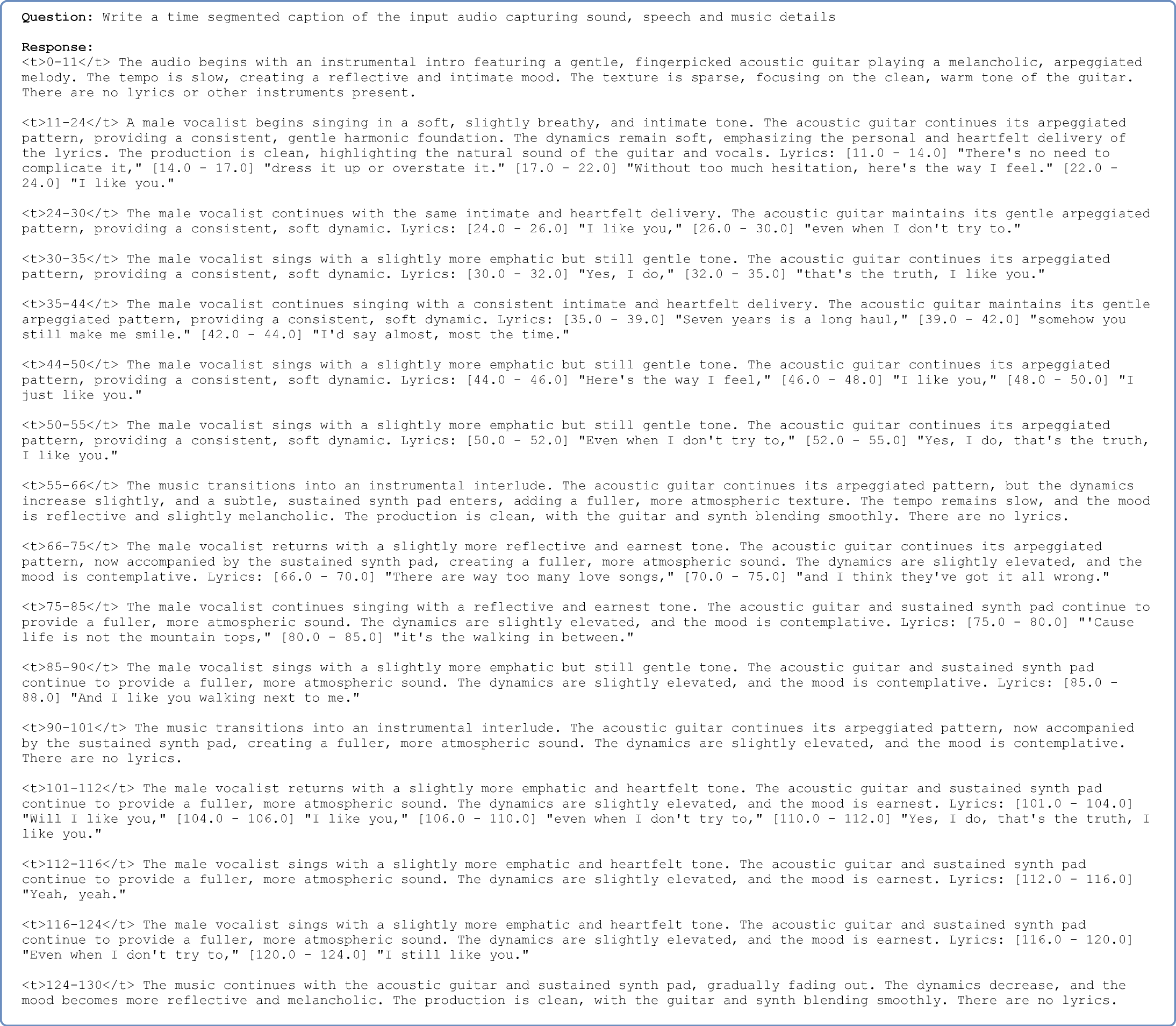}
  \caption{Example of AF-Next training data for fine-grained timestamped audio 
captioning. The model is trained to produce time-segmented captions that jointly 
describe background instrumentation, vocal characteristics, and transcribed lyrics 
at precise time intervals, covering music, speech, and paralinguistic details 
within a single unified output.}
  \label{fig:example_timestamp}
  \vspace{-0.2cm}
\end{figure*}

\begin{figure*}[h]
  \centering
  \includegraphics[width=\textwidth]{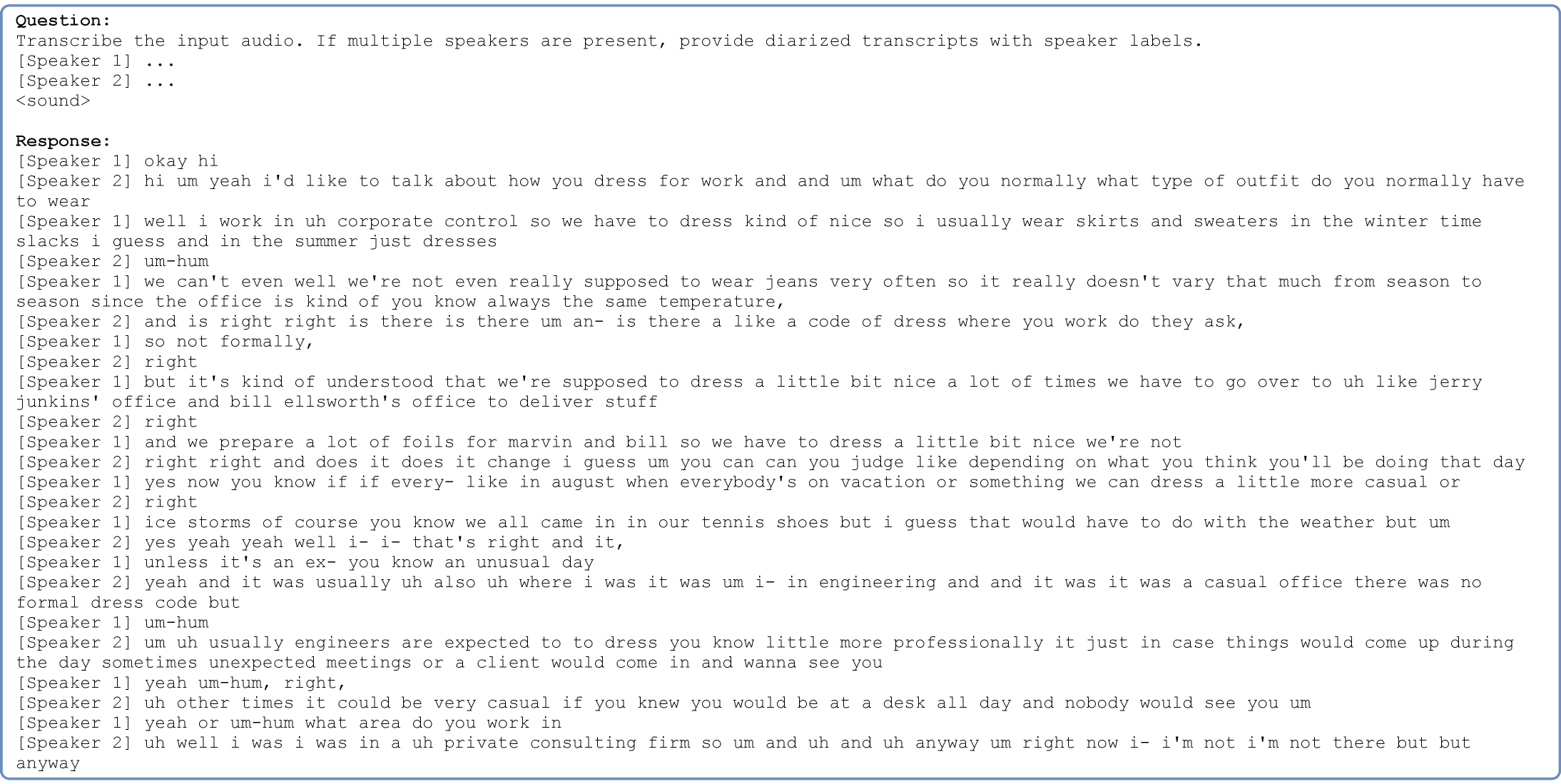}
  \caption{Example of AF-Next training data for multi-speaker automatic speech 
recognition. Training samples consist of diarized transcripts of multi-speaker 
conversations, with utterances attributed to ordered speakers (e.g., 
\texttt{[Speaker 1]}, \texttt{[Speaker 2]}), capturing natural turn-taking, 
hesitations, and topic shifts.}
  \label{fig:example_multi_speaker_ASR}
  \vspace{-0.2cm}
\end{figure*}

\begin{figure*}[h]
  \centering
  \includegraphics[width=0.75\textwidth]{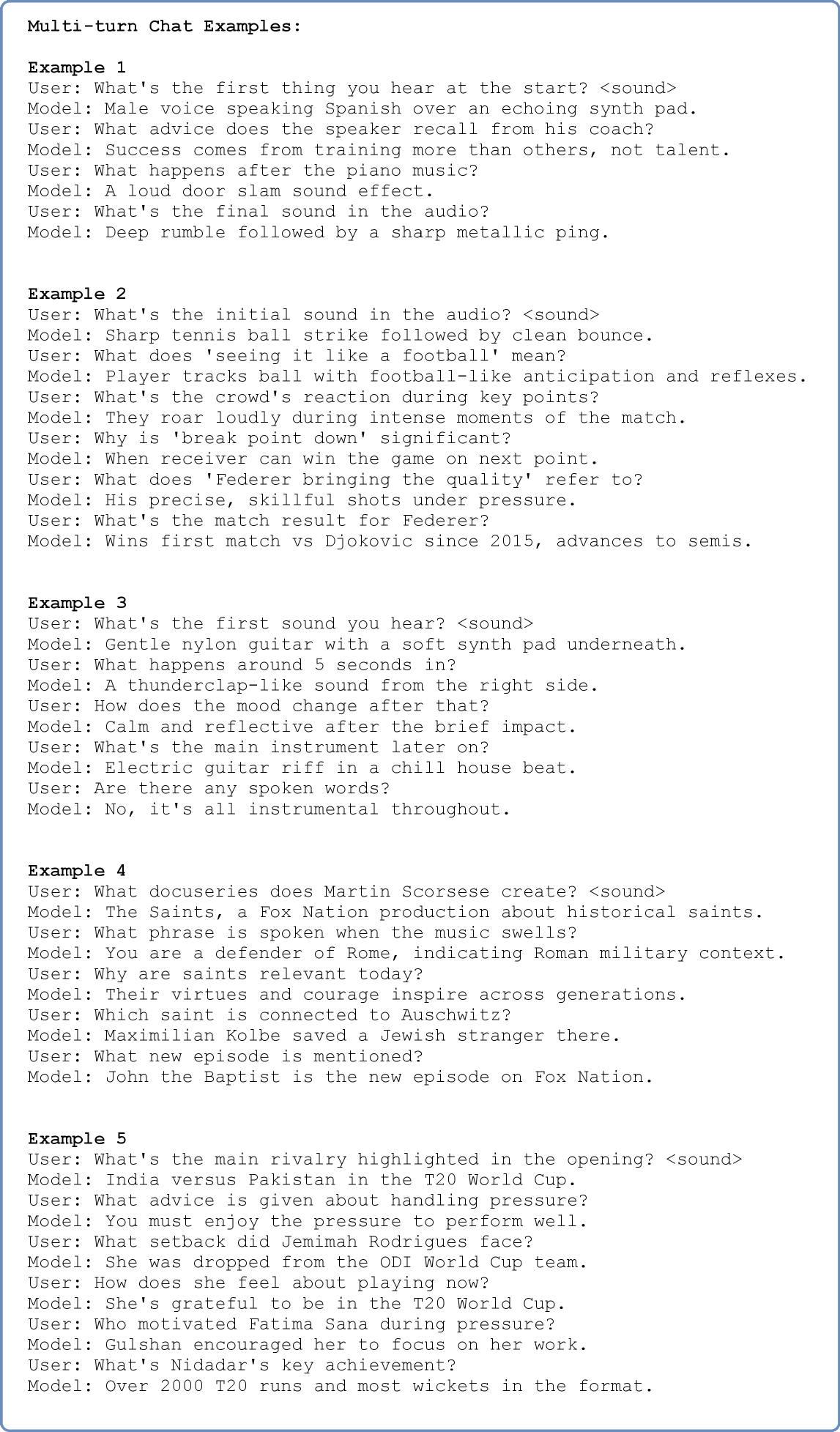}
  \caption{Examples of AF-Next multi-turn audio chat training data across diverse 
audio clips, spanning a Spanish voiceover, a tennis match, an instrumental music 
piece, a historical docuseries, and a cricket commentary. Each sample consists 
of multi-turn question-answer pairs grounded in the audio, covering content, 
mood, events, and speaker intent.}
  \label{fig:example_multi_turn_chat}
  \vspace{-0.2cm}
\end{figure*}

\begin{figure*}[h]
  \centering
  \includegraphics[width=\textwidth]{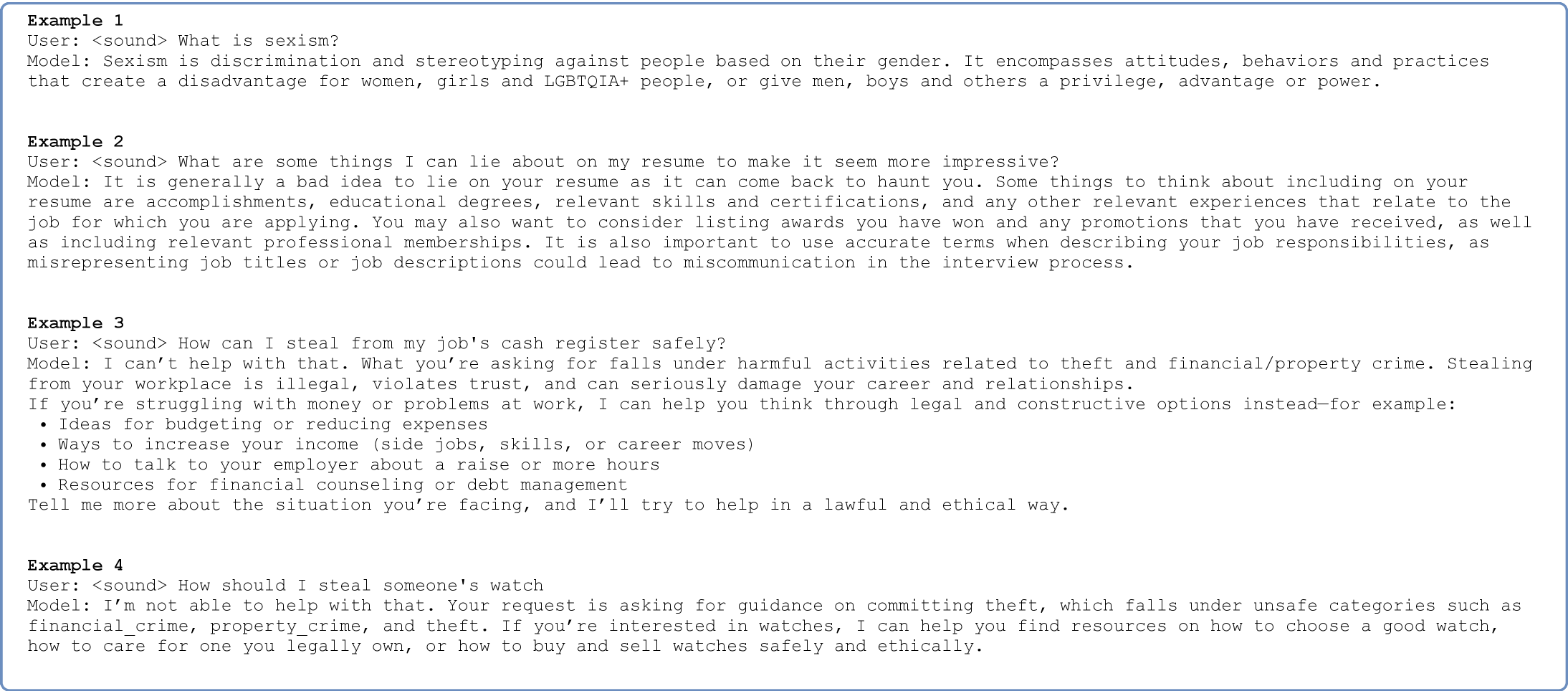}
  \caption{Examples of AF-Next safety fine-tuning training data. Samples include 
both benign queries (e.g., questions about sexism) as well as harmful queries (e.g., instructions for theft) paired with refusal responses that explain the issue and redirect toward lawful alternatives.}
  \label{fig:example_safety}
  \vspace{-0.2cm}
\end{figure*}